\documentclass[aps,pre,reprint,doublespace]{revtex4-1}%
\usepackage{amsmath}%
\usepackage{amsfonts}%
\usepackage{amssymb}%
\usepackage{graphicx}
\usepackage{epstopdf}
\usepackage{subcaption}
\usepackage{setspace}

\begin{document}
\title{Interdependent Lattice Networks in High Dimensions}
\author{Steven Lowinger, Gabriel A. Cwilich, and Sergey V. Buldyrev}
\affiliation{Department of Physics, Yeshiva University, 500 West 185th Street, New
York, New York 10033, USA}

\date{\today}
\begin{abstract}
     We study the mutual percolation of two interdependent lattice
     networks ranging from two to seven dimensions, denoted as $D$. We impose that the
     length of interdependent links connecting nodes in the two
     lattices be less than or equal to a certain value, $r$. For each value of $D$ and $r$, we find the mutual percolation threshold, $p_c[D,r]$ below which the system completely collapses through a cascade of failures following an initial destruction of a fraction $ (1-p)$ of the nodes in one of the lattices. We find that for each
     dimension, $D<6$, there is a value of $r=r_I>1$ such that for
     $r\geq r_I$ the cascading failures occur as a discontinuous
     first order transition, while for $r<r_I$ the system undergoes a
     continuous second order transition, as in the classical
     percolation theory. Remarkably, for $D=6$, $r_I=1$ which is the
     same as in random regular (RR) graphs with the same degree
     (coordination number) of nodes.
     We also find that in all
     dimensions, the interdependent lattices reach maximal
     vulnerability (maximal $p_c[D,r]$) at a distance $r=r_{max}>r_I$, and for $r>r_{max}$
     the vulnerability starts to decrease as $r\to\infty$. However the
     decrease becomes less significant as $D$ increases and $p_c[D,r_{max}]-p_c[D,\infty]$ decreases exponentially with $D$. We also investigate the dependence of $p_c[D,r]$ on the system size as well as how the nature of the
     transition changes as the number of lattice sites, $N\to\infty$.
     \end{abstract} \maketitle
\section{Introduction}
The behavior of many complex systems in the real world can be better
understood and explained through network
theory\cite{Albert2002,Dorogovtsev2003,Satorras2004,Newman2006,Caldarelli2007,
  Newman2010,Cohen2010}. Highway traffic, power outages, the relationship between  businesses  and many other phenomena can be modeled as
networks. Additionally,  many of the real networks, such as the
communications network and the power grid are
interdependent on each other\cite{Buldyrev2010,Parshani2010,Buldyrev2011,Gao2012,Son2011,Son2012,Baxter2012,Peixoto2012,Li,Bashan,korn}. This 
phenomenon can be discussed in terms of  mutual percolation: in order to
function properly, a node in each network must be connected to the
giant component of its network and must be supported by an
interdependent node in the other network. A failure of a fraction $(1-p)$ of nodes in one network
will lead to failures in the other network. This will either cause
both networks to eventually stabilize, preserving their giant components, or to
completely collapse. The communication network and the
power grid network are examples of such interdependent networks, embedded in
space. A blackout in a city may cause a server operating the power
grid to go down, and this may cause further disruption of power
stations. However, it is reasonable to assume that the
interdependent nodes in the two networks are not located far away from each other.  Another
example is the network of seaports and the
network of national highways, which are interdependent on each
other. Hurricane Sandy demonstrated that
if a seaport gets damaged, the city to which it supplies fuel will
become isolated from the highway network. Similarly, a city without fuel for trucks cannot supply a
seaport properly, and the seaport will not be able to function
well \cite{korn}. Often, real world interdependent networks contain nodes which are
embedded in a two dimensional surface or in a three-dimensional
space \cite{Li,Bashan}. Li \textit{et. al} \cite{Li} introduced the concept of
a dependence on distance, according to which a node in network A can be
interdependent with a node in network B, only if the distance
between these two nodes does not exceed a value, $r$. The constraint on the
length of these interdependency links will significantly affect the
mutual percolation of the two networks and will alter the properties
of the system's collapse. They found that for $r=0$, the collapse
transition in  two two-dimensional lattice networks is identical to the classical percolation problem on one
two dimensional lattice\cite{Stauffer1994,Bunde1996}.  As $r$
increases, the critical percolation threshold, $p_c$, increases, but
the transition remains a second order transition. In a second order transition, the
size of the surviving mutual giant component of the system gradually approaches zero as the
fraction $p$ approaches $p_c$.
Interestingly, when $r$ reaches a critical value, $r_I\approx 8$, the
transition suddenly becomes a first order transition, in which either
the majority of the nodes survive, or the networks are completely
destroyed.  As $r$ increases further, $p_c$ starts to decrease until,
for $r\to \infty$, it reaches the value characteristic of the mutual
percolation on the lattices with random interdependency links. In the
interval $r_I\leq r <\infty$, the cascading failures lead to a small
hole which starts to grow circularly until all the nodes of both
lattices are wiped out. The explanation of this phenomena, by Li \textit{et. al} \cite{Li} was based on
the idea that cascading failures in this regime propagate by the destruction of nodes
close to the perimeter of the hole that is larger than $r$. This will happen because such nodes have lost their
supporting nodes in the other network, previously located in the
hole. For small $r$, $p_c$ is close enough to the $p_c$ of classical
percolation, $p_c^p$, at which the size of the holes diverge, so that holes
larger than $r$ appear at the first stage of the cascade. However as
$r$ grows, $p_c$ also grows and eventually the typical size of the holes,
dictated by the correlation length of the classical percolation
becomes equal to $r$. When this happens, the system becomes
metastable: a random formation of a hole of a sufficient size by a
local density fluctuation causes the circular growth of such a hole,
destroying the entire system. As $r$ increases in the vicinity of $r_I$, a
smaller value of  $p$ is needed to produce a hole of size $r$. Therefore, $p_c$
starts to decrease for $r>r_I$. It is clear that all of these interesting
phenomena are related to the presence of the surface of a hole which
is valid only in an object of a finite dimension. 

Indeed, one can more generally study the problem of a propagating  $D-1$ dimensional interface on a
D-dimensional lattice based on the mutual percolation rules discussed above, with the
maximal interdependence distance $r$ and the initial density of surviving sites
$p$. The process of this propagation is similar to the various models
of fluid propagation in disordered media \cite{Barabasi1995} which are
characterized by the depinning transition: i.e. there is a critical
threshold $p=p_c^f$ above which the interface is completely blocked by
the obstacles, but below which the velocity of the interface 
propagation is finite, and  gradually decreases to zero when approaching the critical threshold:  $v\sim (p_c^f-p)^\theta$.  The
depinning transition is a second order transition characterized by
several critical exponents, one of which is  $\theta>0$. The fluid
propagation near $p=p_c^f$ is characterized by avalanches: one
remaining active site in a completely blocked interface can create an
avalanche of propagation. The size distribution of the avalanches obeys
a power law similar to the distribution of the cluster sizes in 
percolation theory. In the mutual percolation model, $p_c^f(r)$ of the
interface propagation increases with $r$ from the value $p_c^p$ of  classical percolation theory
at $r=0$, to the value 1 at $r=r_f$. If $r>r_f$ the interface  propagates freely through
the system even if the lattice is completely intact. 

When $p_c^f$ is close
to the percolation threshold $p_c^p$, the correlation length of percolation, 
$\xi(p_c^f)$, is greater than $r$. This means that there are always holes of size greater than $r$,
and the interface is always spontaneously created. The interface will start to propagate from many
different places. However, if $p$ is close to $p_c^f$ from above, the propagation will stop
leaving a sponge-like mutual giant component with holes of all possible
sizes. The death of a single node may disconnect a huge portion of the mutual 
giant component and may dramatically reduce its size. Hence there is a broad
distribution of the sizes of the mutual giant component, which is one of the  
characteristic of a second order phase transition. In contrast, when $p_c^f$ is
far above $p_c$, for large values of $r$, the size of the holes is smaller than $r$ and the interface cannot be created spontaneously;  so,  one must reduce $p$ in order
for a hole of size $r$ to be created.  As the value of $r$ considered is larger, the value of $p$ required to create such a hole decreases. Once the hole is created, its interface starts
to propagate freely because $p<p_c^f$, and it will wipe out the entire lattice.
 In this scenario,  for small $r$ the critical threshold of the mutual percolation is $p_c=p_c^f(r)$, which increases with $r$, until $\xi(p_c^f)=r$. In that interval, the transition is second order. But, for $r>\xi(p_c^f)$, $p_c$ starts
to decrease following the equation $\xi(p_c)=r$ and the transition becomes first order. In fact, the values of $r=r_I$ at which the transition becomes the 
first order and $r=r_{max}$ at which $p_c$ starts to decrease may not 
exactly coincide. There is always a probability that a hole of size $r>\xi(p)$ may appear
in a large enough system. Thus for $r>r_I$, one can expect the average $p_c$ to be
in between the increasing function $p_c^f(r)$ and the decreasing function $p(r)$ defined by the equation $\xi[p(r)]=r$, and hence may still increase until it reaches its maximum at $r=r_{max}$.
One would expect that the value of $r_{max}$ should depend on the system size. 
         
In a random-regular (RR) graph of degree $k$, which can be regarded as an infinitely
dimensional lattice, the surface is not a well defined concept,
because its dimensionality is equal to the dimensionality of the
entire graph.  Thus, one can hypothesize that for two interdependent
identical RR networks, in which the notion of distance is defined as the shortest path
between a pair of nodes, $p_c$ should monotonically increase with
$r$, and the transition should  become a first order transition for very small
$r$. Indeed, the work of Kornbluth {\it et al.}\cite{korn} confirmed
this hypothesis, showing that for $k>8$,  $r_I=1$. For the case ($k=8$, $r=1$), a
first order phase transition is closely followed by a second order
phase transition at a smaller $p$,  and for $k\leq 7$, $r_I=2$. One can
expect that when the dimensionality of the lattice increases, the
behavior observed for the
$D$-dimensional interdependent lattices studied here must converge to the behavior
of the interdependent RR graphs with the  possible existence of an upper
critical dimension\cite{Stauffer1994,Bunde1996} above which the fractal
dimension of the percolation cluster and the fractal dimension of its
surface (accessible perimeter) both become equal to 4 and, hence, the propagation of an interface 
becomes ill-defined.  For classical percolation the upper critical dimension is
known to be six\cite{Coniglio1984}. Thus one can hypothesize that for
the mutual percolation of  two 6D lattice networks, the behavior will be
similar to that of infinite dimensional networks, for which the interface of a percolation cluster coincides with the cluster itself. The goal of this paper is to test all of the hypotheses discussed above.
         
\section{Model}
We study the mutual percolation\cite{Buldyrev2010} of two interdependent hypercubic lattice networks in several dimensions. We create two identical networks A and B, whose nodes are labeled $1,2,...N=L^D$ where $D$ is the dimension of the lattice and $L$ is the number of nodes along each of its dimensions. Each node is connected with edges to exactly $k=2D$ nearest-neighbor nodes. We then introduce one-to-one bidirectional interdependency links, such that the shortest path between any two interdependent nodes is not greater than $r$. 
In order to decrease computation time and define how the network is built, we introduce two isomorphisms between networks A and B. These isomorphisms, the  topological isomorphism, $\mathcal{T}$, and the dependency isomorphism, $\mathcal{D}$, are those which were defined in Kornbluth, \textit{et. al} \cite{korn}. The topological isomorphism is defined for each node $A_i$ as $\mathcal{T}(A_i)=B_i$ and verifies that if  $A_i$ and $A_j$ are first neighbors in network A,  then $\mathcal{T}(A_i)$ and  $\mathcal{T}(A_j)$  are first neighbors in network B and vice versa. For the case of lattices, the topological isomorphism is automatically established due to  the identical lattice structure. The dependency isomorphism, establishes the interdependency links, and  we create it following the restriction that $B_k=\mathcal{D}(A_i)$ only if the shortest path connecting $A_i$ and $A_k=\mathcal{T}(B_k)$ is of a length, $r_{ik}\leq r$. Since our goal is to compare the behavior of $D$-dimensional hypercubic lattices to the RR graphs with $k=2D$, for which the concept of coordinates is not applicable, we choose our definition of distance as one that is identical to that used for RR graphs (i.e. the smallest number of edges connecting the two sites). In the context of hypercubes this metric is the Manhattan metric, which slightly differs from both the Euclidian metric and the cubic metric used in Li \textit{et al.}\cite{Li},  $r={\rm max}_{i=1}^D{|\Delta x_i|}$, where $\Delta x_i$ are the coordinate differences of the two interdependent nodes.    

In order to establish the dependency isomorphism, while still satisfying the shortest path restriction, Li {\it et al.}\cite{Li} created a random permutation of the indices of all the nodes that fulfilled the distance restriction. However, in our case we followed the procedure developed by Kornbluth {\it et al.}\cite{korn},namely we set $\mathcal{D}(A_i)= B_i$ only if there are no other possibilities for $\mathcal{D}(A_i)$. Additionally, we require that if $\mathcal{D}(A_i)= B_k$, then $\mathcal{D}(B_i)= A_k$. This further restriction decreases the time required for computation, without affecting the results in any essential way.

Initially, a fraction $(1-p)$ of randomly selected nodes in the first network are destroyed. Any node in the second network whose interdependent node in the first network has been destroyed, or who lost its connectivity to the largest percolation cluster (the largest group of nodes, connected to each other) will also be destroyed. We return to the first network and further destroy all the nodes who lost their support in the previous process, or who got disconnected  from the largest percolation cluster, as a consequence of the previous stage. This process of destruction  continues to alternate between the networks and is referred to as a cascade of failures. The process ends when both networks no longer contain nodes that  will fail. The largest cluster of nodes which spans the entire network is called the mutual giant component. In all cases, if the fraction of nodes $p$ surviving the initial attack falls below a certain critical threshold, $p$-critical or $p_c$, the network completely collapses and the giant component disappears.  We study how  this value changes as a function of the maximum distance of interdependent links $r$, as well as the dimensionality of the networks $D$. We denote the $p$-critical value for a network of dimension, $D$, and distance, $r$, as $p_c[D,r]$. We run our simulations for lattices of $N \approx 10^6$ nodes regardless of dimension, unless otherwise specified.

\section{Simulation Results}
\begin{figure}
\includegraphics[clip,width=\columnwidth, height=5cm]{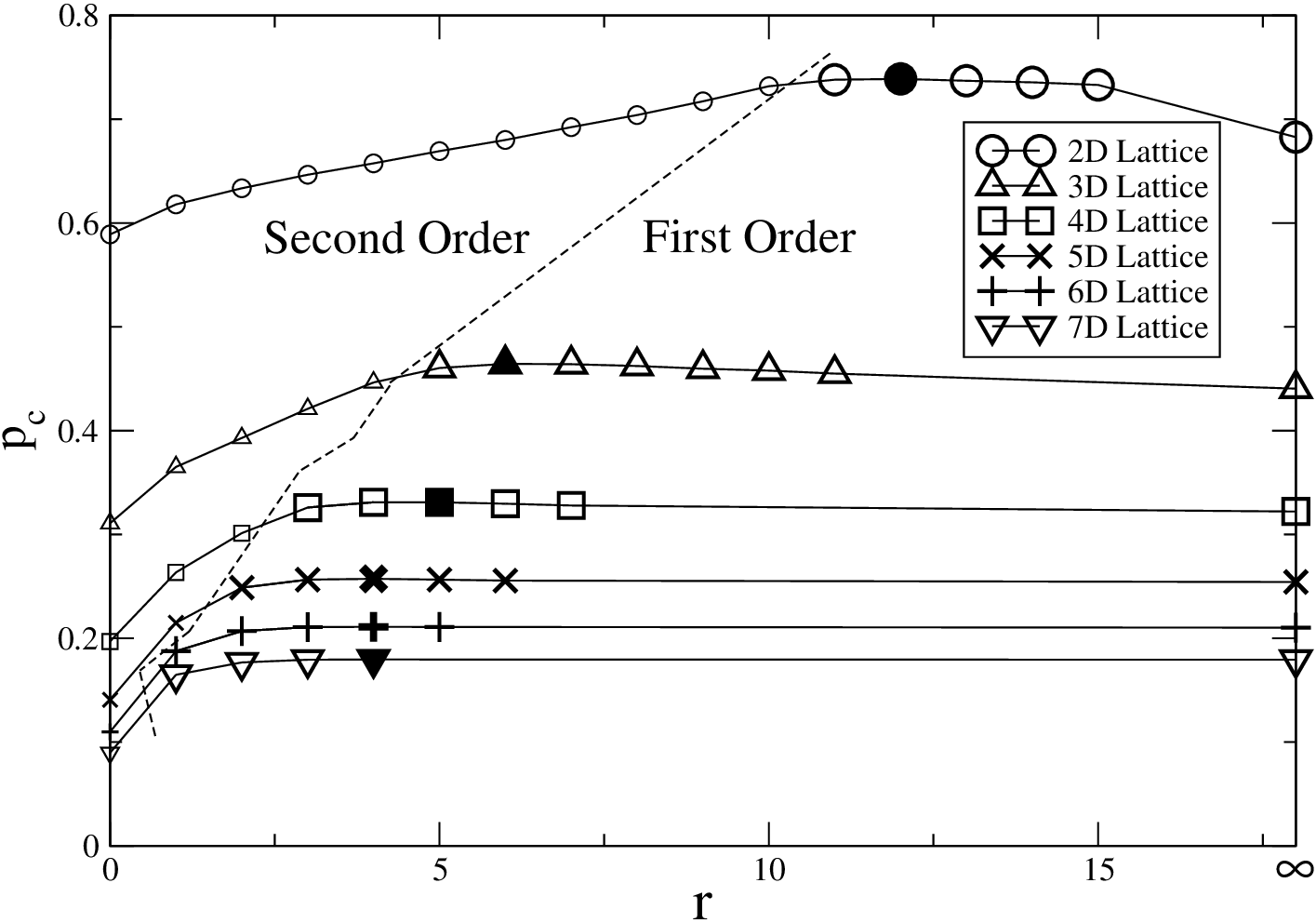}
\caption{Plot of $p_c[D,r]$ vs. $r$ for lattices of dimensions ranging from 2 to 7. The smaller symbols correspond to second order transitions, the larger symbols correspond to first order transitions and the bold symbols denote the maximum value of $p_c[D,r]$ for a given dimension. The last value in each plot is the value of $p_c[D,\infty]$}
\label{f:apc}
\end{figure}

We run simulations to determine the value of $p_c$ for lattices of two through seven dimensions. For these lattices, we find that the value of $p_c$ increases with $r$, reaches a maximum at $r=r_{max}$, and then slowly converges to $p_c[D,\infty]$, which is the value of $p_c$ for random interdependency links (Figure \ref{f:apc}). For low values of $r<r_I$, the transition is second order, while for higher values of $r\geq r_I$ the transition is first order. Additionally, we find that $r_{max}>r_I$ for all $D$. For example, in a two-dimensional lattice for $0 \leq r \leq 10$, the transition is second order and for $r \geq r_I=11$ the transition is first order and the maximum value of $p_c$ occurs when $r=r_{max}=12$.\\

\begin{figure}
\includegraphics[clip,width=\columnwidth, height=5cm]{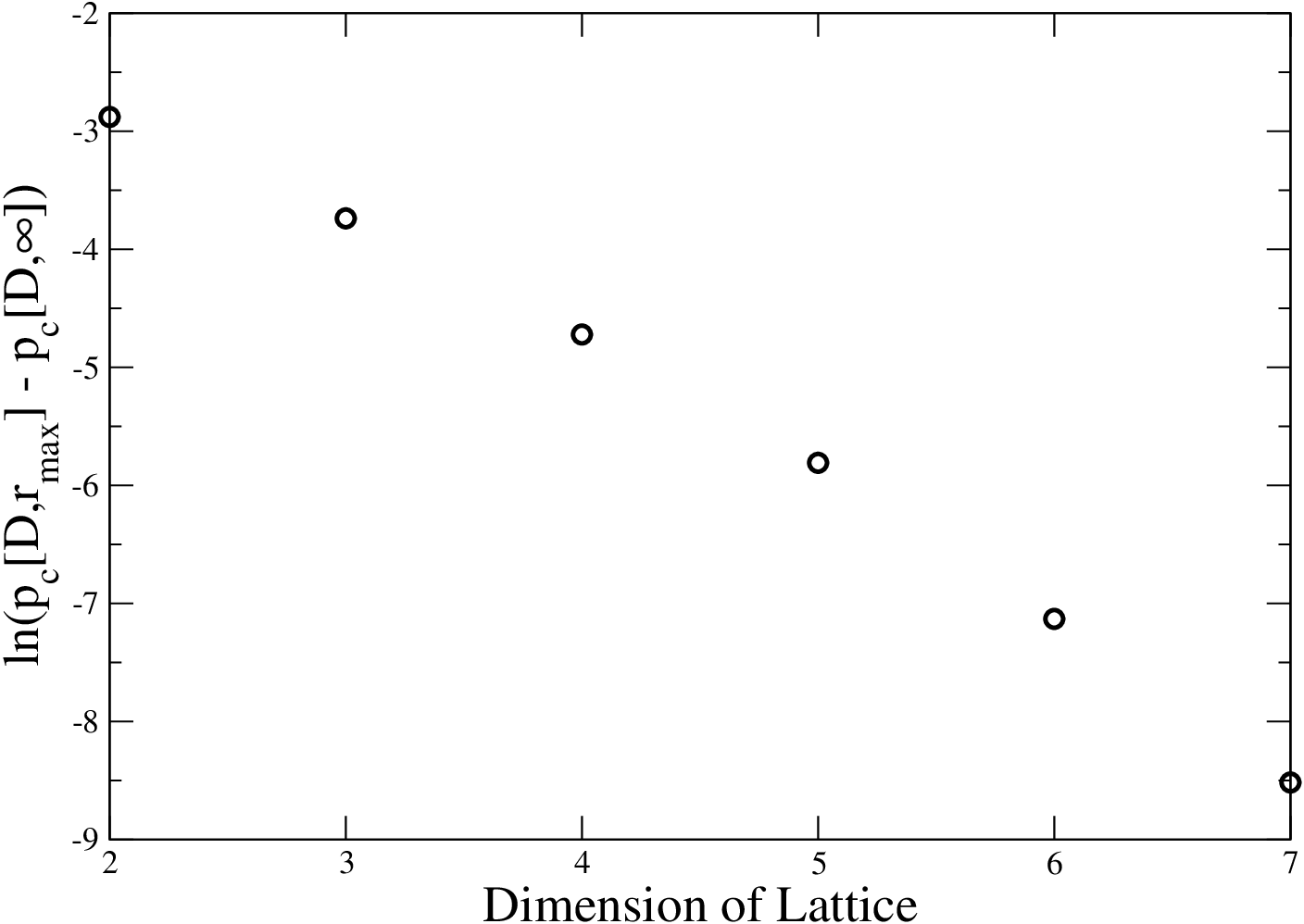}
\caption{Plot of the decrease of the difference between $p_c[D, r_{max}]$ and $p_c[D,\infty]$ as dimension increases.}
\label{f:maxinf}
\end{figure}

The trend of the maximum value of $p_c$ occurring after the change from second to first order transitions is present through all dimensions, including the seven dimensional lattice. However, the difference, $p_c[D,r_{max}]-p_c[D,\infty]$, decreases exponentially with $D$ (Fig. \ref{f:maxinf}). It is also interesting that the difference between $r_{max}$ and $r_I$ increases with $D$. Thus the case of RR graphs in which the maximum of $p_c$ is reached for $r=\infty$ \cite{korn}, is the limiting case of the behavior of lattices when $D\to\infty$.\\

\begin{figure*} 
\begin{subfigure}[b]{.32\textwidth}
\centering
\includegraphics[clip,width=\textwidth]{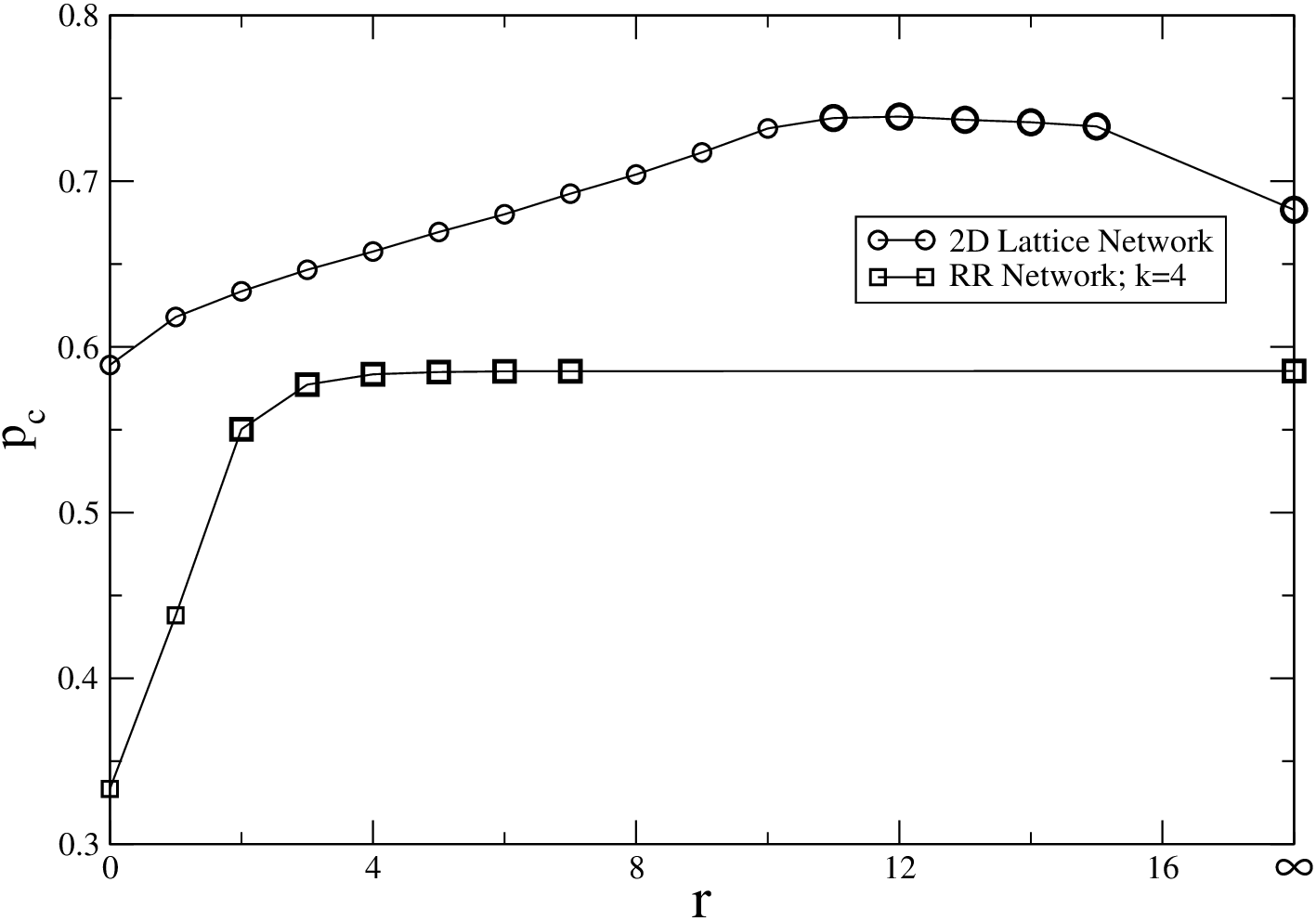}\hfill
\caption{2D lattice and RR network with $k=4$}
\label{f:2dcmp}
\end{subfigure}
\quad
\begin{subfigure}[b]{.32\textwidth}
\centering
\includegraphics[clip,width=\textwidth]{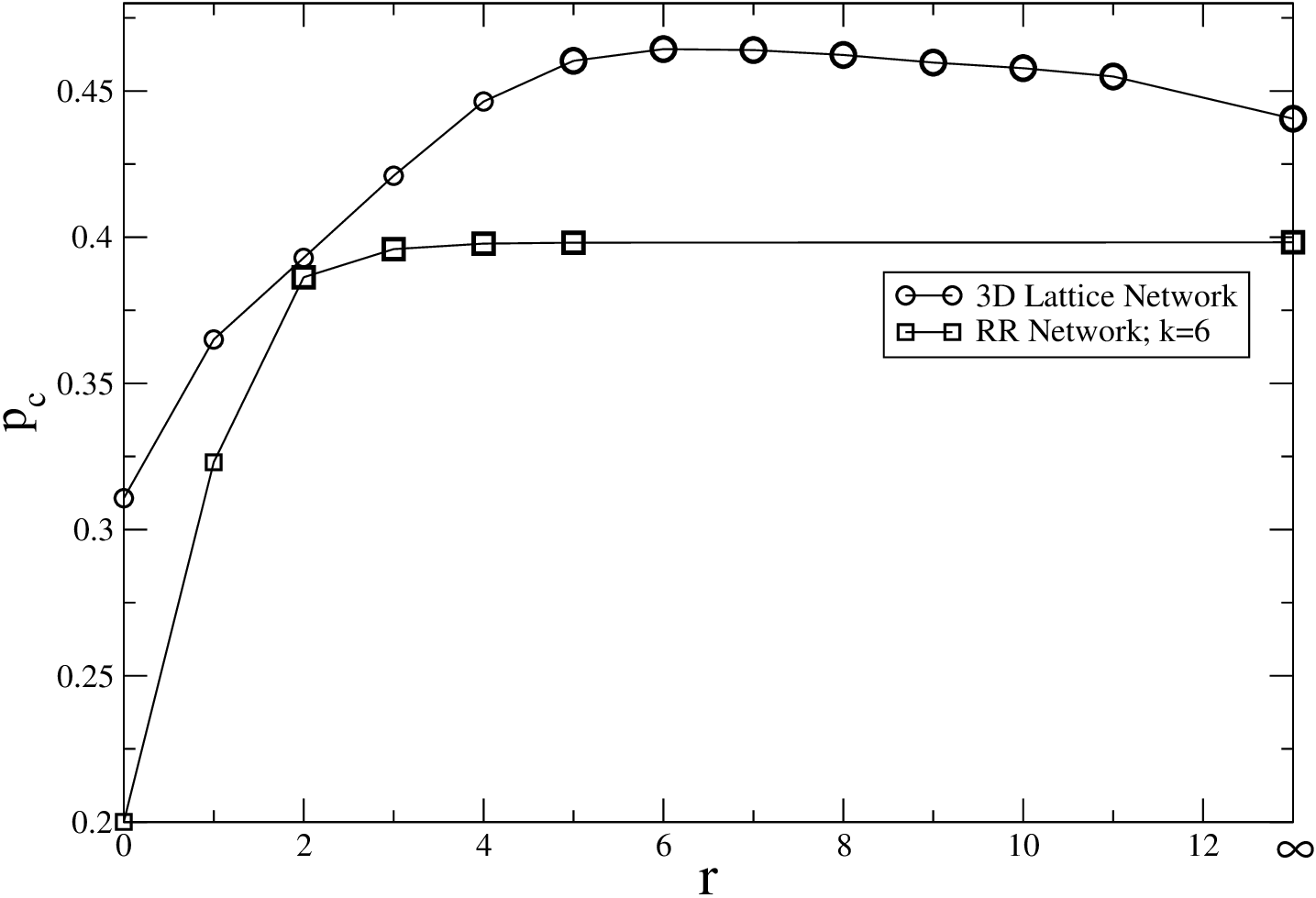}\hfill
\caption{3D lattice and RR network with $k=6$}
\label{f:3dcmp}
\end{subfigure}
\begin{subfigure}[b]{.32\linewidth}
\centering
\includegraphics[clip,width=\linewidth]{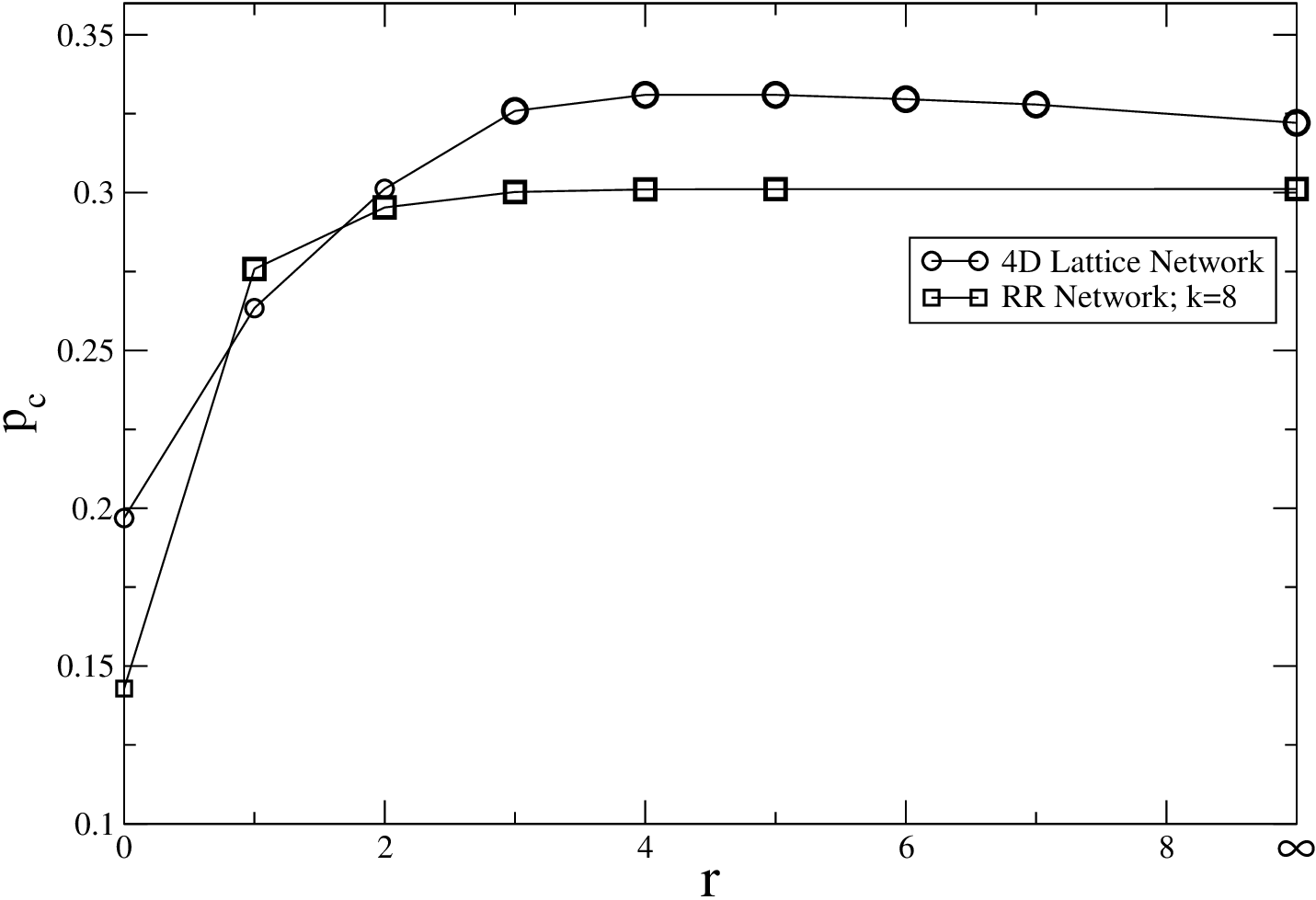}\hfill
\caption{4D lattice and RR network with $k=8$}
\label{f:4dcmp}
\end{subfigure}
\\
\begin{subfigure}[b]{.32\linewidth}
\centering
\includegraphics[clip,width=\linewidth]{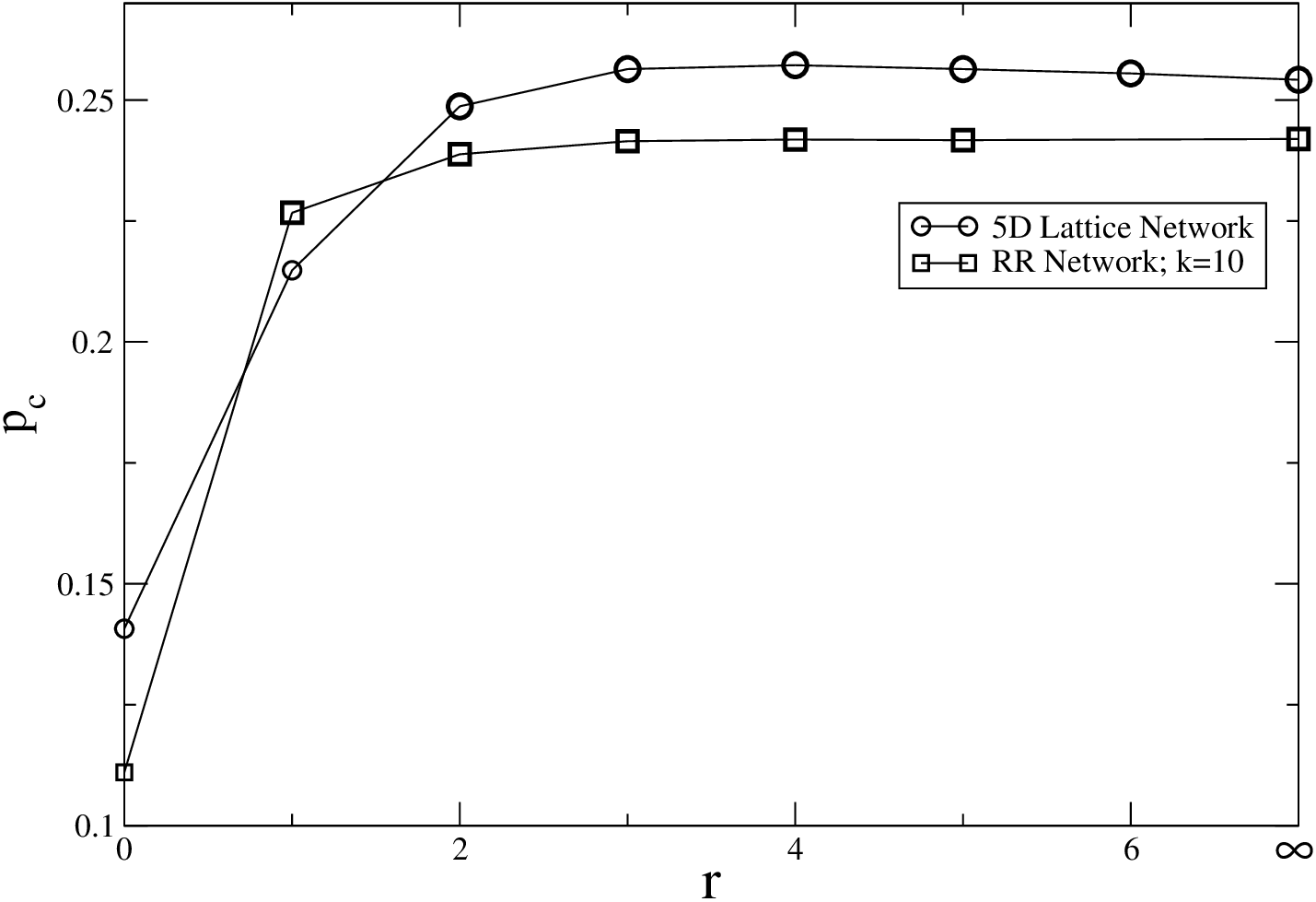}
\caption{5D lattice and RR network with $k=10$}
\label{f:5dcmp}
\end{subfigure}
\quad
\begin{subfigure}[b]{.32\linewidth}
\centering
\includegraphics[clip,width=\linewidth]{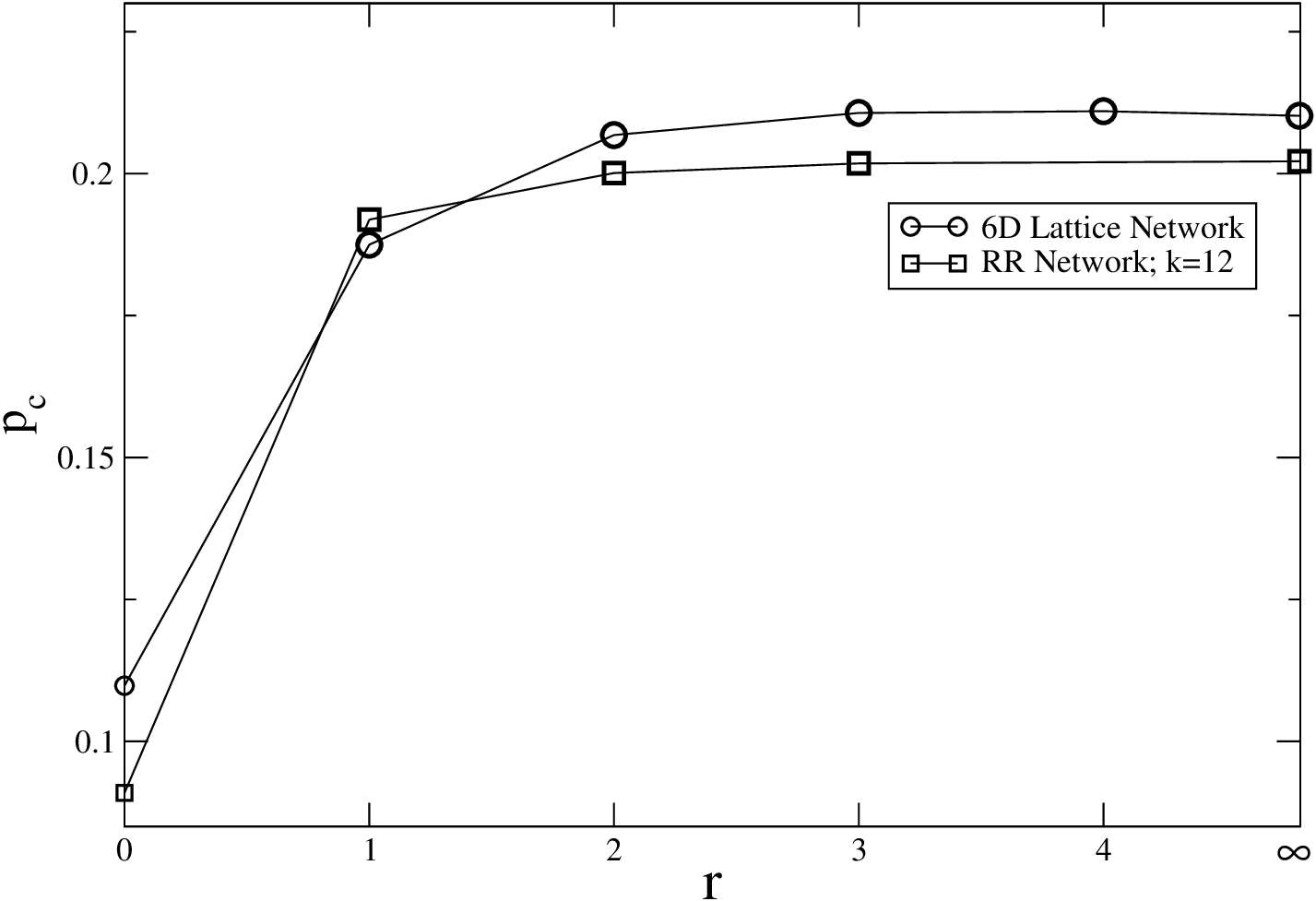}
\caption{6D lattice and RR network with $k=12$}
\label{f:6dcmp}
\end{subfigure}
\begin{subfigure}[b]{.32\linewidth}
\centering
\includegraphics[clip,width=\linewidth]{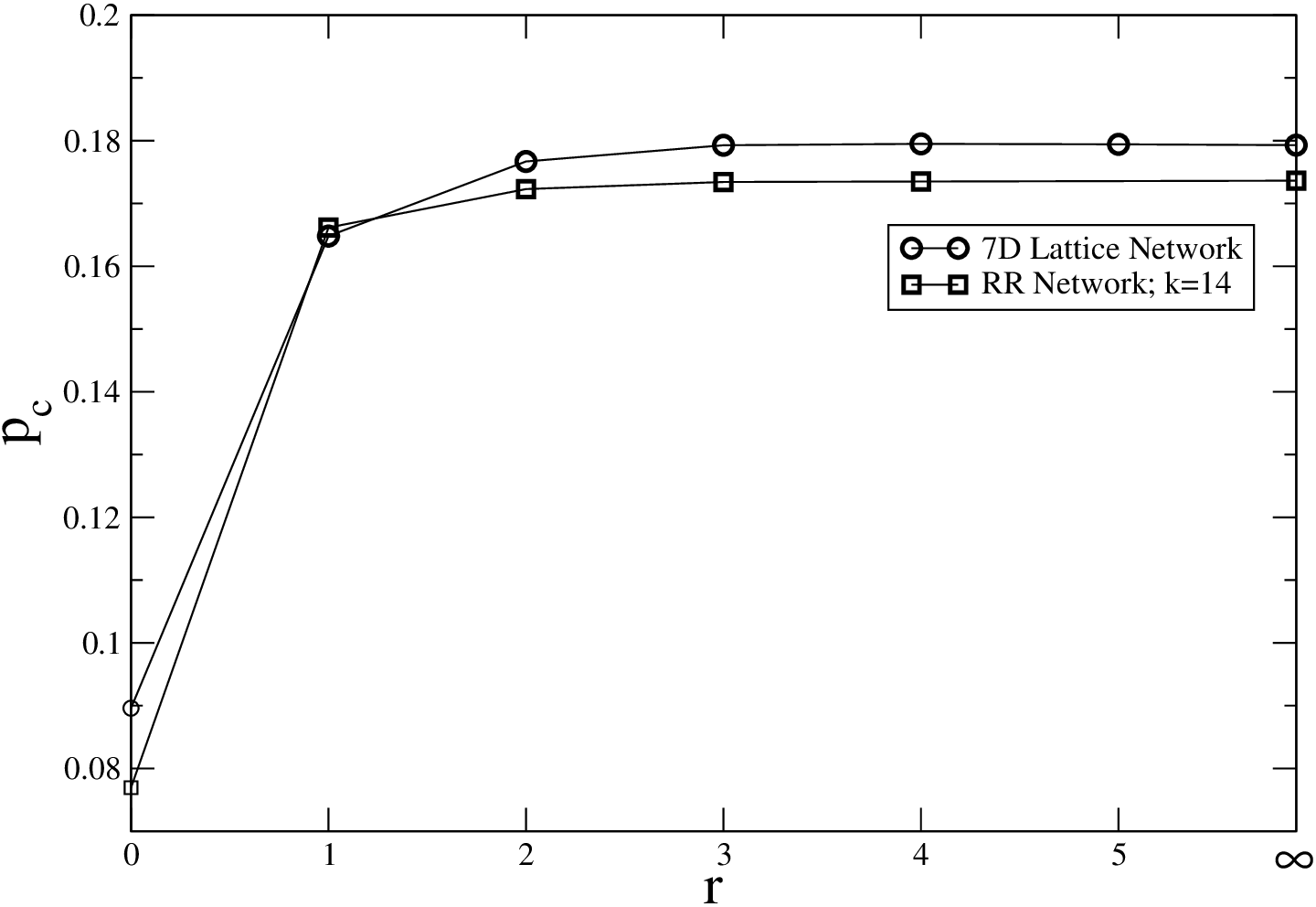}
\caption{7D lattice and RR network with $k=14$}
\label{f:7dcmp}
\end{subfigure}
\\
\caption{Comparison of $p_c[D,r]$ vs. $r$ for lattices of varying dimensions and the corresponding RR network. The smaller symbols denote second order transitions and the larger symbols denote first order transitions.}
\label{f:ndcmp}
\end{figure*}

\begin{figure}
\includegraphics[clip,width=\columnwidth, height=5cm]{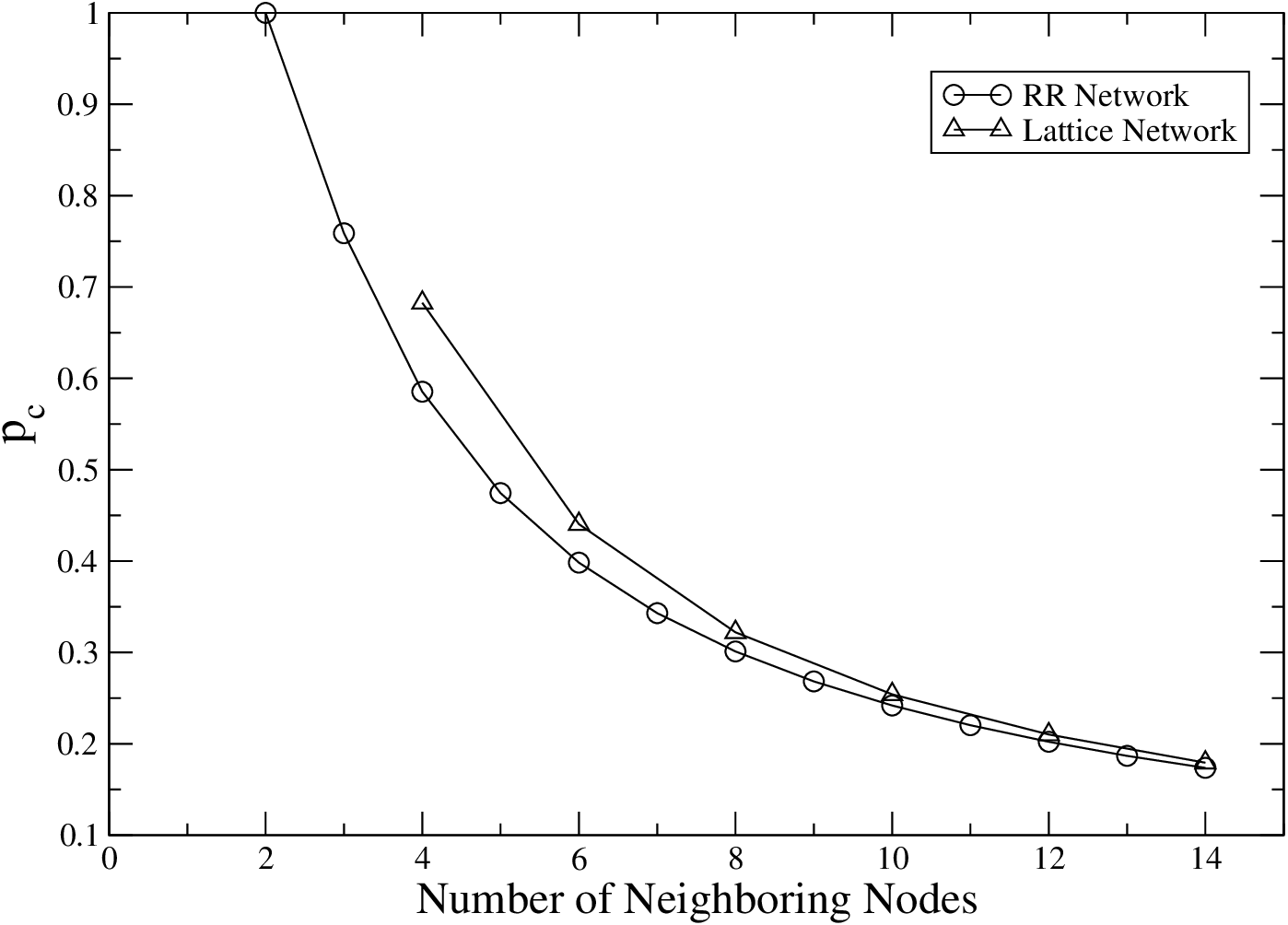}
\caption{Comparison of our simulation results for $p_c$ of lattice networks to random interdependent links as well as the analytical results for $p_c$ for RR networks with $k=2D$ and random interdependent links}
\label{f:inf}
\end{figure}

Additionally, as $D$ increases, the difference between the individual values of $p_c$ for the lattice and RR networks with $k=2D$, decreases (Fig. \ref{f:ndcmp}). Figure \ref{f:inf} shows a comparison between the $p_c[D,\infty]$ for our simulation results and the analytical results for $p_c$ for a RR network with $k=2D$ and random interdependent links. The $p_c$ of the lattice network slowly approaches that of the RR network as the number of neighboring nodes (degree), $k$, increases.

\section{Types of Collapses}
\begin{figure}
\includegraphics[clip,width=\columnwidth]{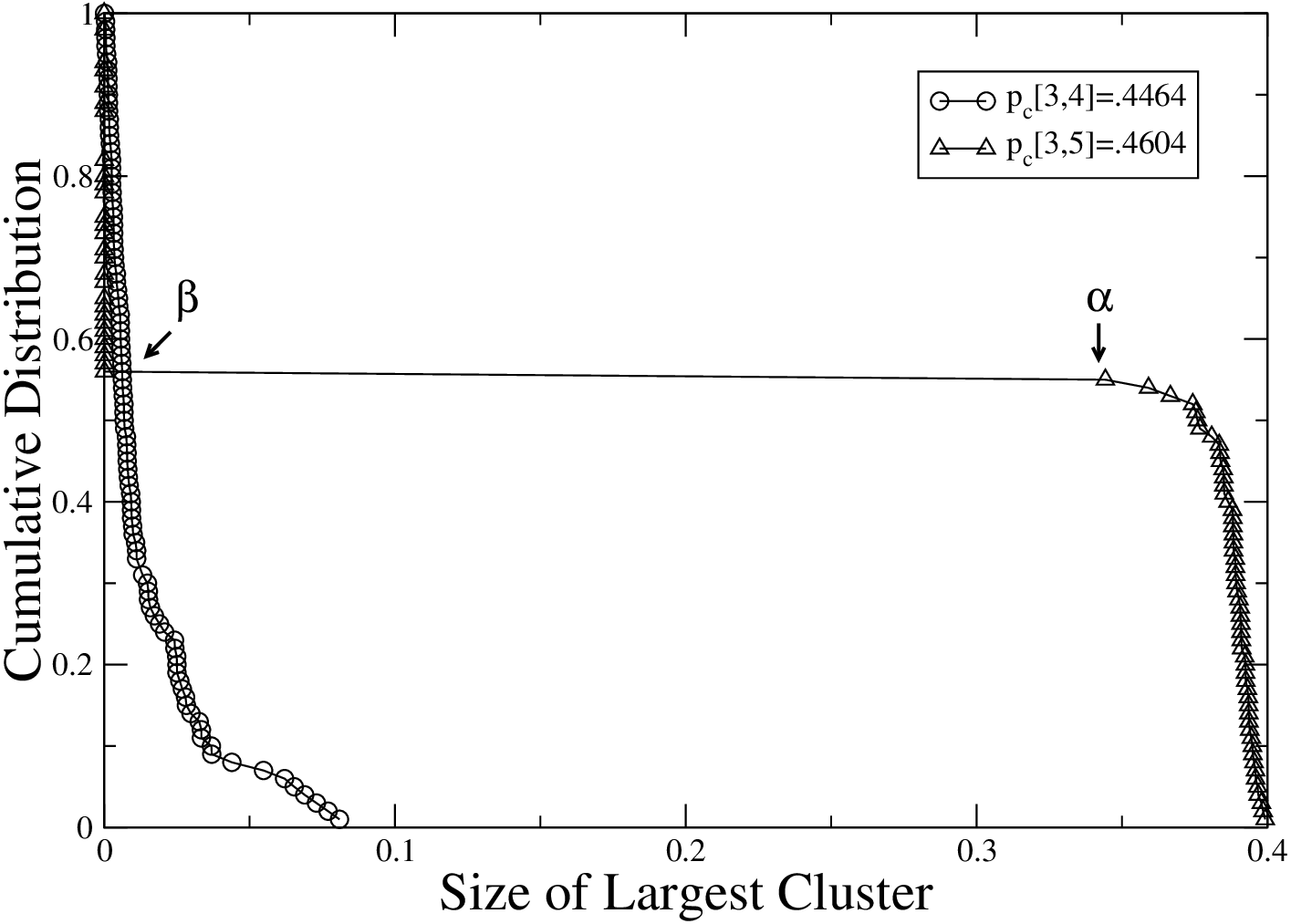}
\caption{Plot of the cumulative distribution of the largest cluster for the last second order transition of a 3D network with $N=10^6$, at $p_c[3,4]=.4464$ and the first first order transition at $p_c[3,5]=.4604$. There is an obvious large gap in the plot of the size of the largest cluster for the first order transition.}
\label{f:FO}
\end{figure}

\begin{figure}
\includegraphics[clip,width=\columnwidth, height=5cm]{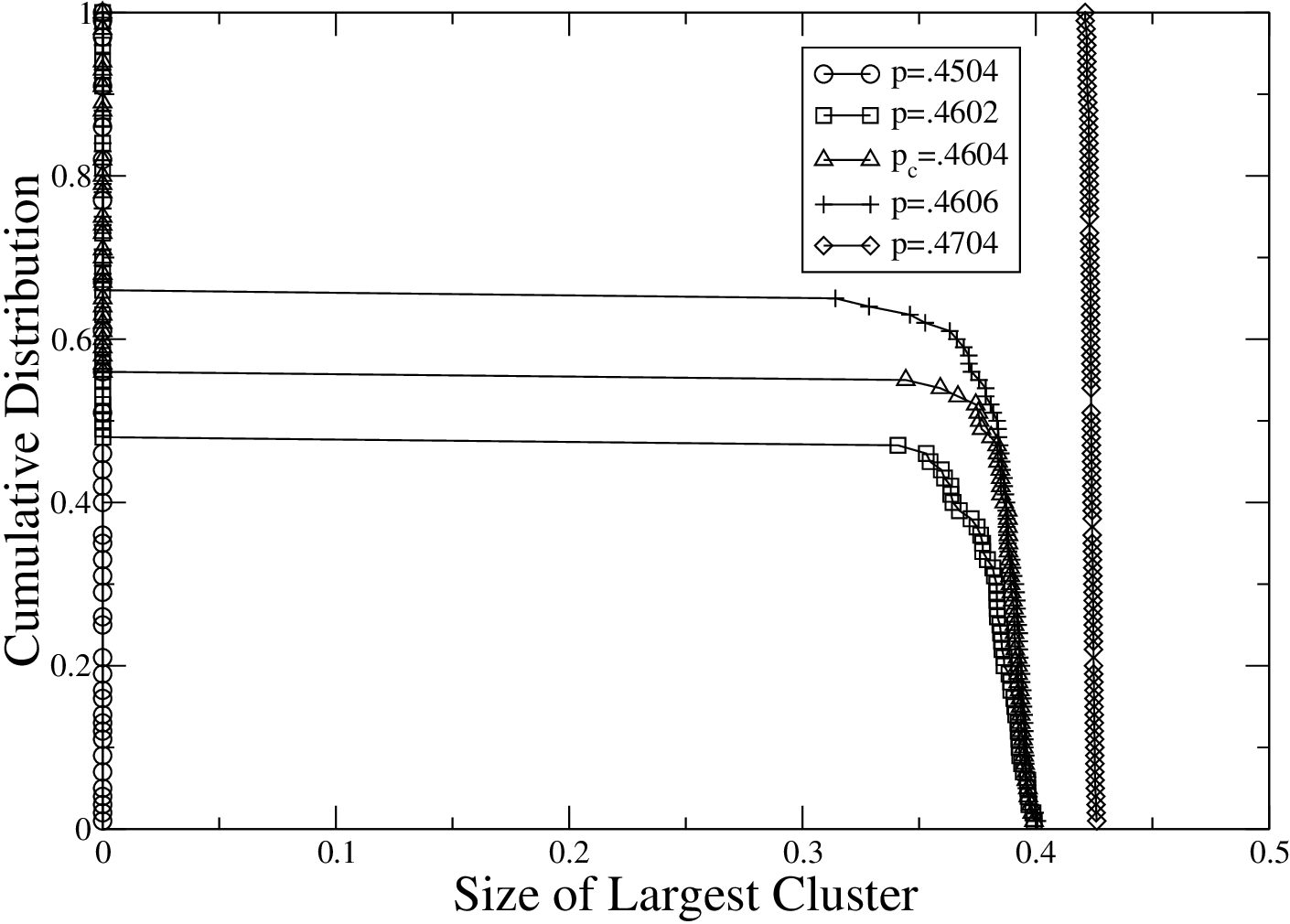}
\caption{Plot of largest cluster size at and near $p_c[3,5]=.4604$. It can be seen that a very small change in $p$ leads to drastically different largest cluster size distributions.}
\label{f:FOC}
\end{figure}

We find that for a $D$-dimensional lattice network, the network experiences two different types of collapses: a first order transition and a second order transition. There are various characteristics which differ between these transitions. One such difference is in the graph of the cumulative distribution of the mutual giant component of the networks at $p=p_c$. After a network undergoes a first order transition, there is a single giant component of size $\mu N$ as well as many small clusters of sizes significantly less than $\mu N$. For finite $N$ we define the giant component as the largest cluster. As seen in Fig. \ref{f:FO}, for a first order transition there are boundaries on the possible size of the giant cluster. We refer to the last resulting cluster size preceding the gap (looking from the left), as $\beta$, and the first cluster size following the gap as $\alpha$. There are no simulations which result with $\alpha > \mu > \beta$. 
This phenomenon tells us that if a giant component becomes smaller than $\alpha$, then it will not become stable until it falls below $\beta$. For networks of infinite size, for $p>p_c$, the fraction of nodes in the mutual giant component is equal to $\alpha$ and for $p<p_c$ it is equal to 0. In finite networks, there is always an uncertainty in the size of the giant component and we define $p_c$ as the point at which approximately half of the realizations lead to largest clusters larger than $\alpha$ and approximately half smaller than $\beta$. This discontinuity in the distribution of the size of the largest cluster characterizes the first order transition of interdependent networks, and occurs due to a cascade of failures. Another feature of a first order transition is the dramatic change in the giant cluster distribution, with a small change in $p$. As seen in Fig. \ref{f:FO}, exactly at $p_c$, approximately half of the realizations result with largest cluster smaller than $\beta$ and approximately half larger than $\alpha$. However, the plots of the cumulative distribution of the largest cluster for values of $p$ slightly larger or smaller than $p_c$ look significantly different (Fig. \ref{f:FOC}).
For a second order transition the graph of the cumulative distribution of the mutual giant component of the networks looks significantly different than that of a first order transition. During the collapse, there is a slow decline in the size of the network, and the size of the giant component can take many values with no discontinuous jump in the middle of the distribution (Fig. \ref{f:FO}).

\begin{figure}
\begin{subfigure}[b]{.45\columnwidth}
\centering
\includegraphics[clip,width=\columnwidth]{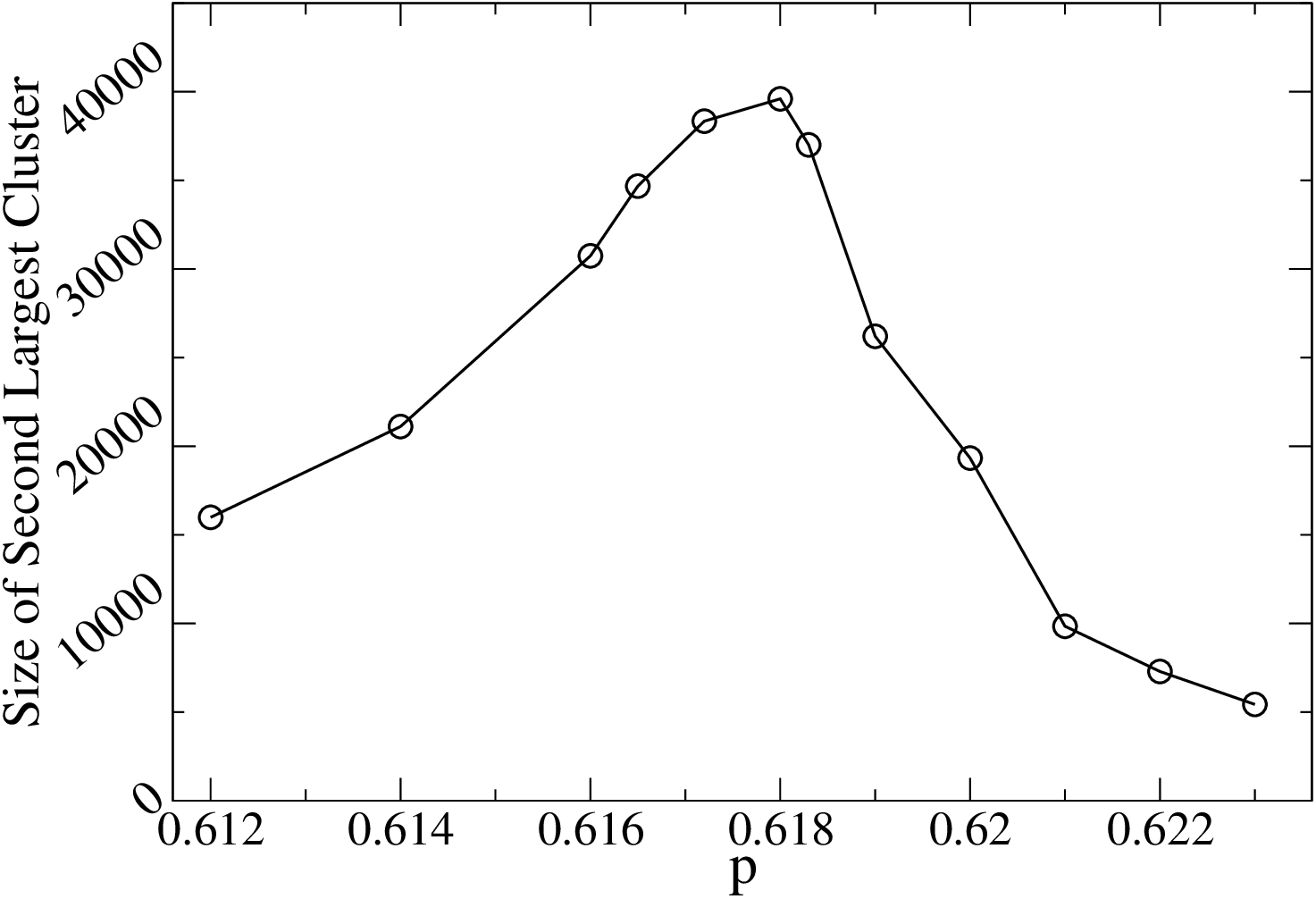}
\caption{2D Lattice with $p_c[2,1]=.618$}
\label{f:slc2}
\end{subfigure}
\quad
\begin{subfigure}[b]{.45\columnwidth}
\centering
\includegraphics[clip,width=\columnwidth]{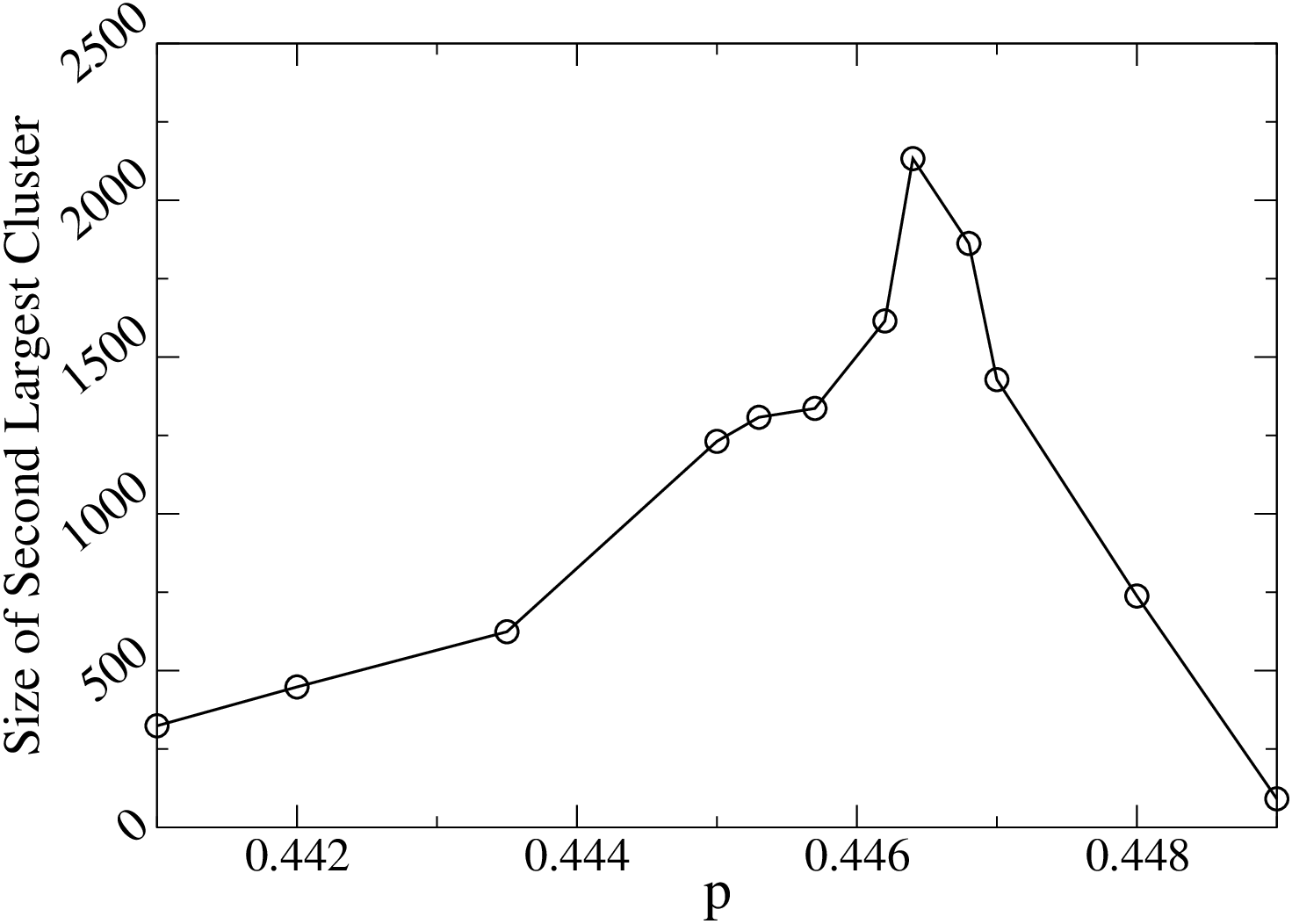}
\caption{3D Lattice with $p_c[3,4]=.4464$}
\label{f:slc3}
\end{subfigure}
\begin{subfigure}[b]{.45\columnwidth}
\centering
\includegraphics[clip,width=\columnwidth]{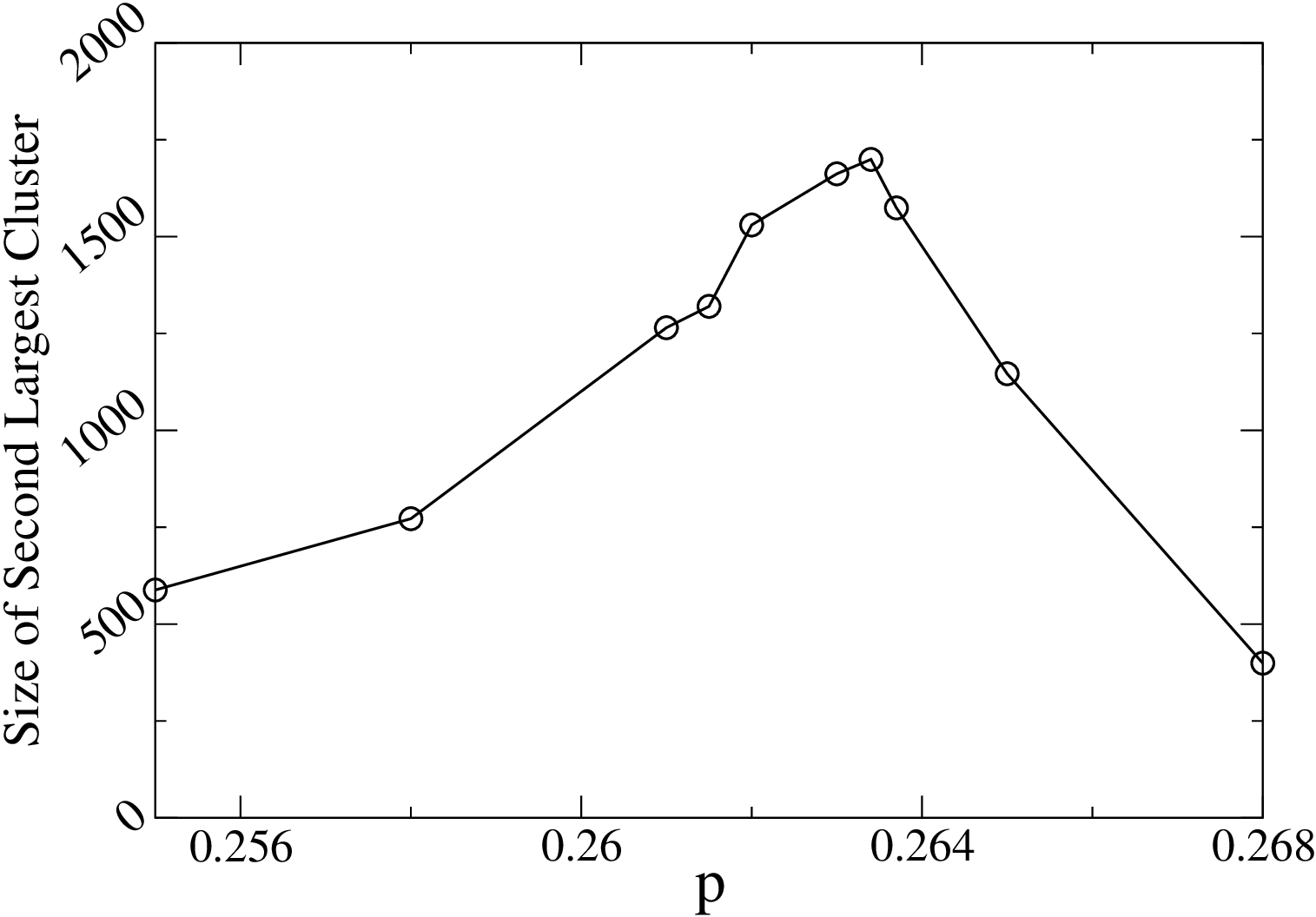}
\caption{4D Lattice with $p_c[4,1]=.2634$}
\label{f:slc4}
\end{subfigure}
\quad
\caption{Plot of the average second largest cluster size as a function of $p$ for 2, 3 and 4 dimensional lattices}
\label{f:slc}
\end{figure}

In order to determine $p_c$ and determine if a network undergoes a first or second order transition, we examine the graphs of the distribution of the largest cluster as well as the distribution of the second largest cluster. First we study the plot of the largest cluster distribution. If we determine that the distribution is characteristic of a first order transition, we define $p_c$ as the value of $p$ such that approximately half of of the simulations produce robust networks and half produce completely collapsed networks. However, if the distribution is reminiscent of a second order transition, we use the distribution of the second largest cluster to determine $p_c$. As discussed in Kornbluth \textit{et al.} \cite{korn}, when $p>p_c$ the giant cluster spans the majority of the network, preventing other large clusters from forming. When $p<p_c$ the network is very broken-up, and large clusters are, therefore, not able to form. However, when $p=p_c$, the average size of the second largest cluster reaches a sharp peak (Fig. \ref{f:slc}). This trend is present in the second order transitions of all of the lattices, regardless of dimension. Thus, for the case of second order transitions, we define $p_c$ as the value of $p$ for which the average size of the second largest cluster reaches its maximum.

The order of any transition is only well-defined in the thermodynamic limit of $N\to \infty$. Therefore, in order to properly model networks of infinite size, one must generate a network that is sufficiently large. The network size sufficient for the correct determination of the order of the transition varies for networks of different dimensions. If the network is too small, then finite size effects will affect the network's behavior as well as the type of transition. In our simulations, we find that, in general, $N=10^6$ is a sufficient number of nodes to model networks of infinite size and the transitions represent the true transitions of networks of infinite size for that dimension. However, we notice that if the distance $r$ immediately follows or precedes the change from second order transition to first order transition, for a given $D$, the system exhibits the strongest finite size effects.

\begin{figure}
\includegraphics[clip,width=\columnwidth, height=5cm]{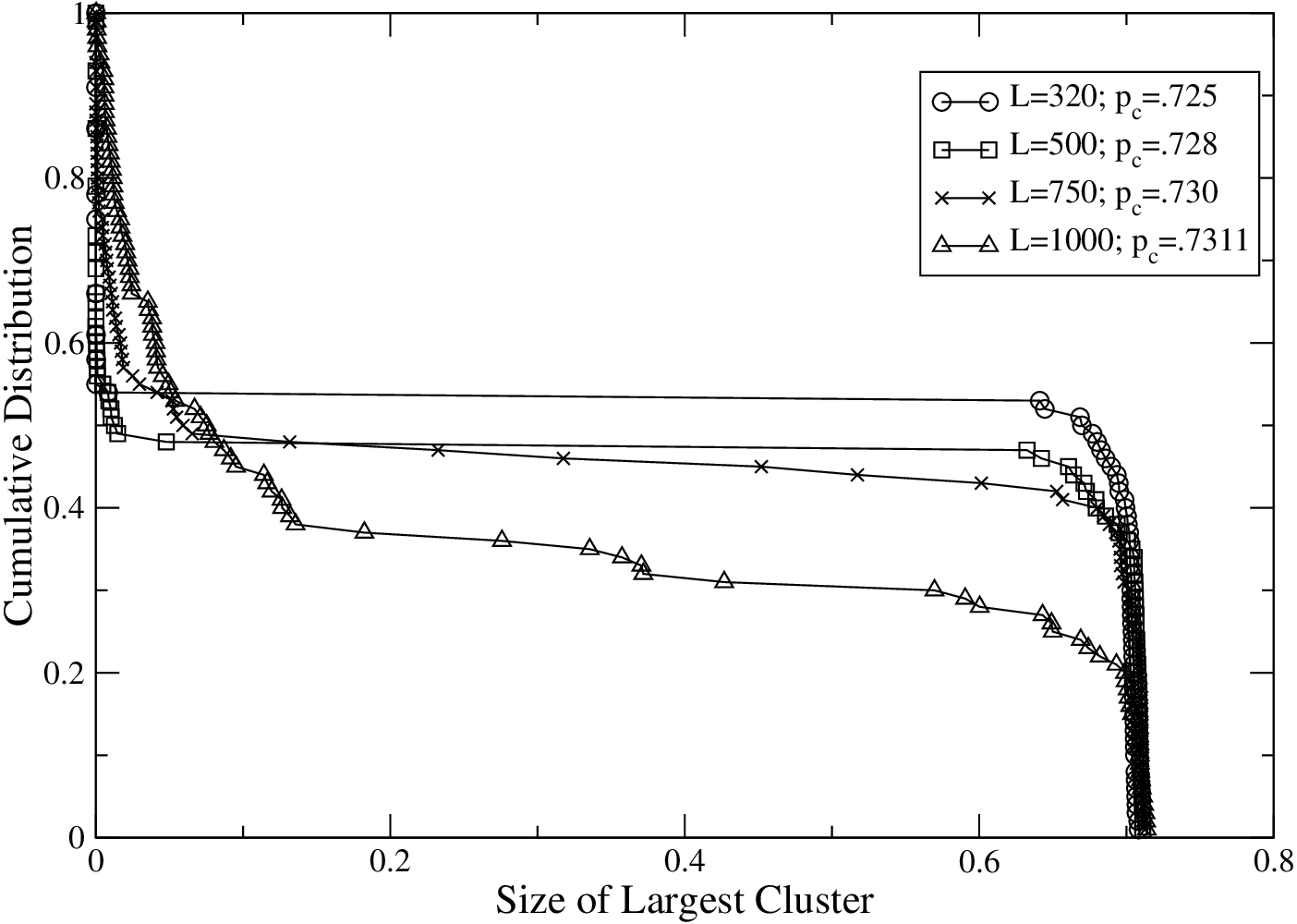}
\caption{Plot of the largest cluster distribution for a 2D lattice network with $r=10$ of increasing size. It can be seen that as the size of the network increases, the type of transition becomes more second order. When $L\geq 750$, the transition becomes second order and approaches the true transition of the 2D lattice.}
\label{f:Lpc2}
\end{figure}

\begin{figure}
\includegraphics[clip,width=\columnwidth, height=5cm]{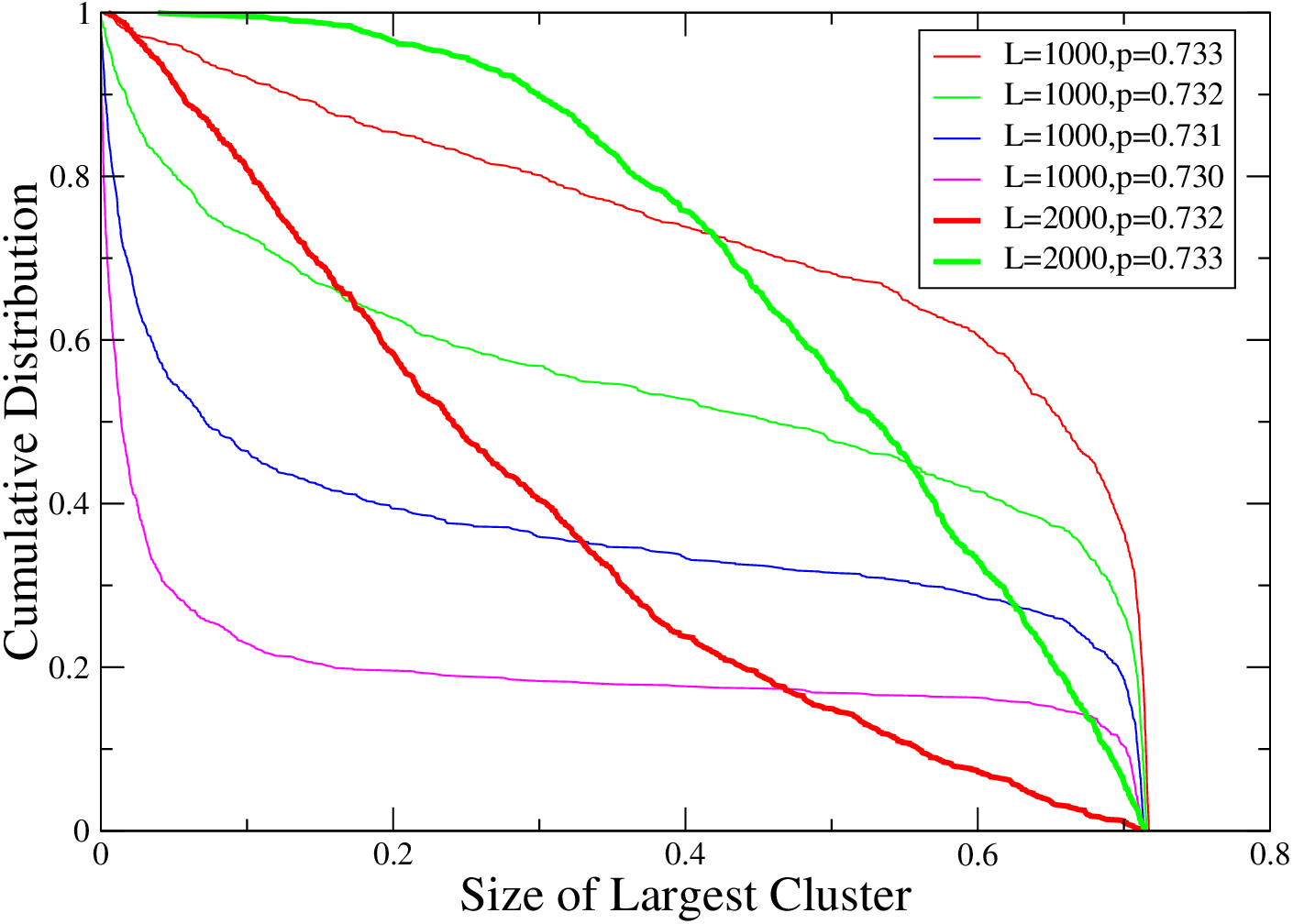}
\caption{(Color online) Plot of the largest cluster distribution for a 2D lattice network of size $L=1000$ and $L=2000$, with $r=10$, for different values of $p$. When $L=2000$ the transition is completely second order.}
\label{f:2-10}
\end{figure}

\begin{figure}
\includegraphics[clip,width=\columnwidth, height=5cm]{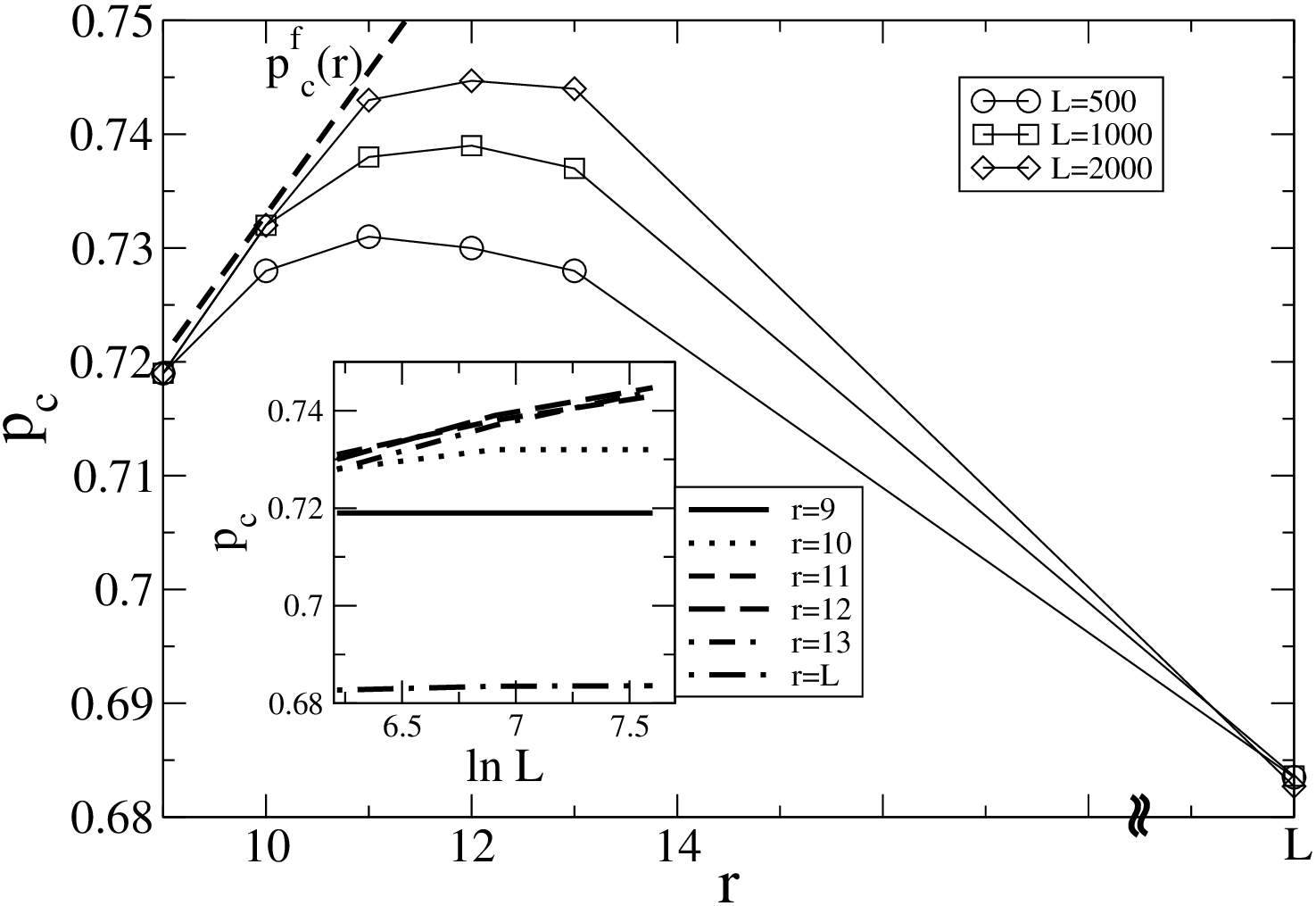}
\caption{Behavior of $p_c$ for $D=2$, as function of $r$ for different system sizes $L=500,1000$,and $2000$. Inset shows $p_c$ as function of $\ln(L)$ to test Eq. (\ref{e:p}).}
\label{f:Pc2N}
\end{figure}

\begin{figure}
\includegraphics[clip,width=\columnwidth, height=5cm]{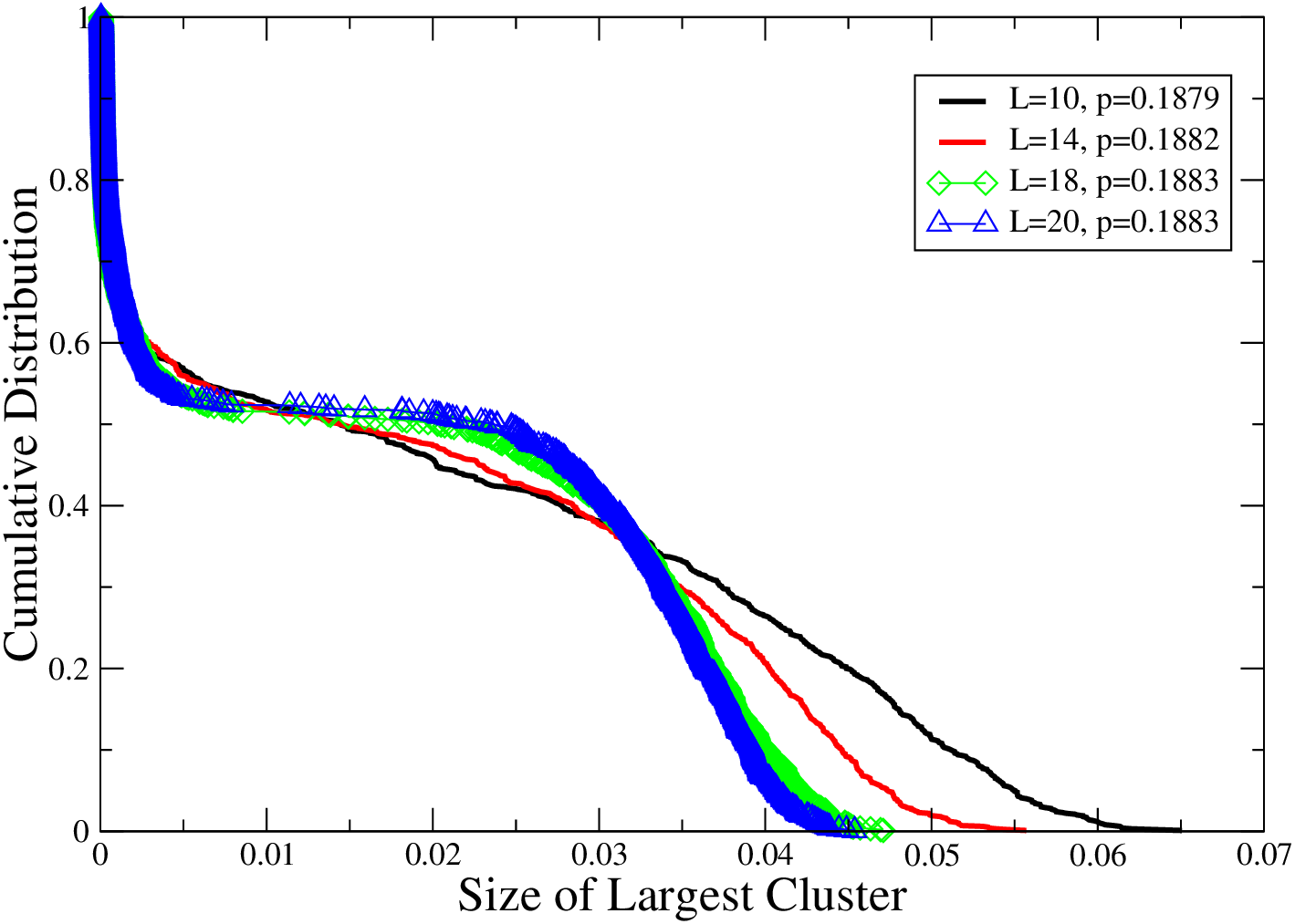}
\caption{(Color online) Plot of the largest cluster distribution for a 6D lattice network with $r=1$ of increasing size. It can be seen that as the size of the network increases, the type of transition becomes more and more first-order-like. When $L=20$, the transition becomes completely first order and the finite size effects are no longer present.}
\label{f:Lpc}
\end{figure}

In all cases studied we find that change of the order of the transition only occurs in the cases of $D=2$, $r=10$ (Figs.~\ref{f:Lpc2} and \ref{f:2-10}) and $D=6$ and $r=1$ (Fig.~\ref{f:Lpc}). These two examples are very different from each other and the changes of the order of the transition are caused by different mechanisms. The formation of the large holes as the mechanism of the network collapse is especially important in low dimensional systems in which the dimensionality of the interior of the hole and its perimeter are significantly different. As mentioned in the introduction, the critical threshold of the free interface moving, $p_c^f(r)$, linearly increases with $r$. However, the probability of the spontaneous formation of a hole of size $r$ at $p_c^f(r)$, decreases with $p_c^f$. This is because the probability, per one lattice site, of the formation of a hole of size $r$ decreases exponentially with $r$, $p_h(r)\sim\exp(-r/\xi)$ where $\xi\sim(p_c^f-p^p_c)^{-\nu}$ is the percolation correlation length, $p^p_c$ is the percolation critical threshold and $\nu$ is a critical exponent. The total probability of the formation of a hole in a lattice of size $N$ is $Np_h(r)$ and we can expect the formation of the hole in a given instance of the lattice if $Np_h(r)=1$.
Thus the fraction, $p_h$ of survived nodes for which the hole of size $r$ will be formed can be found from the following equations:
\begin{equation}
\begin{aligned}
L^D\exp[-ar(p_h-p^p_c)^{\nu}]=1\\ 
ar(p_h-p^p_c)^{\nu}=D\ln(L)\\
p_h(r)=(D\ln(L)/ar)^{1/\nu}+p^p_c
\end{aligned} 
\label{e:p}
\end{equation}
where $a$ is a proportionality coefficient. If $p_h<p_c^f(r)$, then the system is metastable and the hole of size $r$ certainly eliminates the entire system. If $p_h>p_c^f(r)$ then the interface of the hole will grow unpredictably, as in the second order phase transition. Thus if $L$ is large enough we can expect the transition to be second order (Fig. \ref{f:Lpc2} and \ref{f:2-10}) and follow the increasing function, $p_c=p_c^f(r)$ for larger and larger $r$ (Fig.~\ref{f:Pc2N}). For a fixed $L$, as soon as $p_h(r)<p_c^f(r)$ the transition will change to the first order. Moreover, $p_c$ will switch to follow the graph of $p_h(r)$. Thus at a fixed $r$, $p_c$ will increase logarithmically with the system size until it reaches $p_c^f(r)$, after which the dependence on $L$ stops
(Fig.~\ref{f:Pc2N}).

\begin{figure}
\includegraphics[clip,width=\columnwidth, height=5cm]{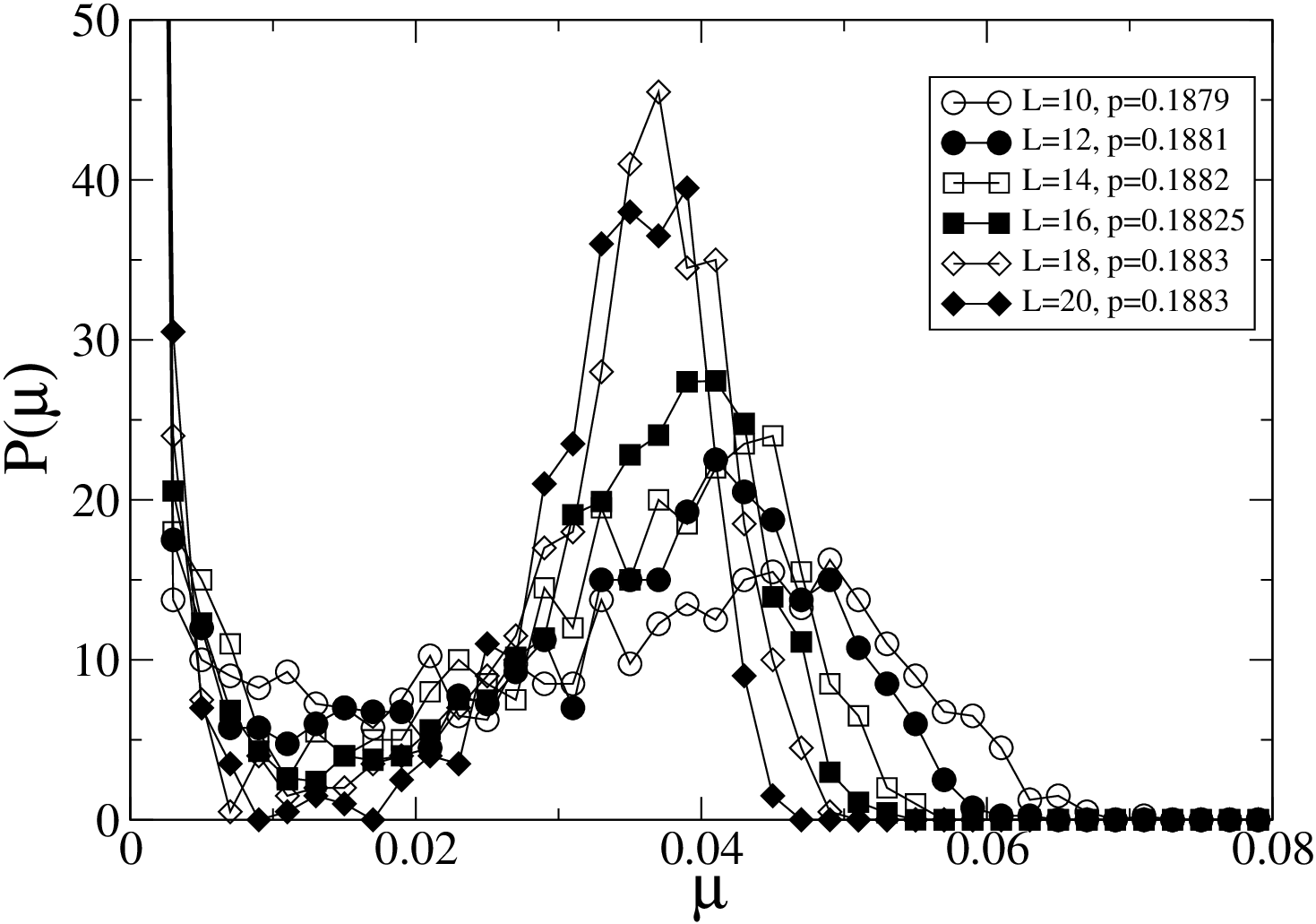}
\caption{PDF of $\mu$ for $D=6$, $r=1$ for increasing values of $L$ from 10 to 20. One can see that the right peak, corresponding to survived giant component, becomes sharper as $L$ increases.}
\label{f:bimodal}
\end{figure}

\begin{figure}
\includegraphics[clip,width=\columnwidth, height=5cm]{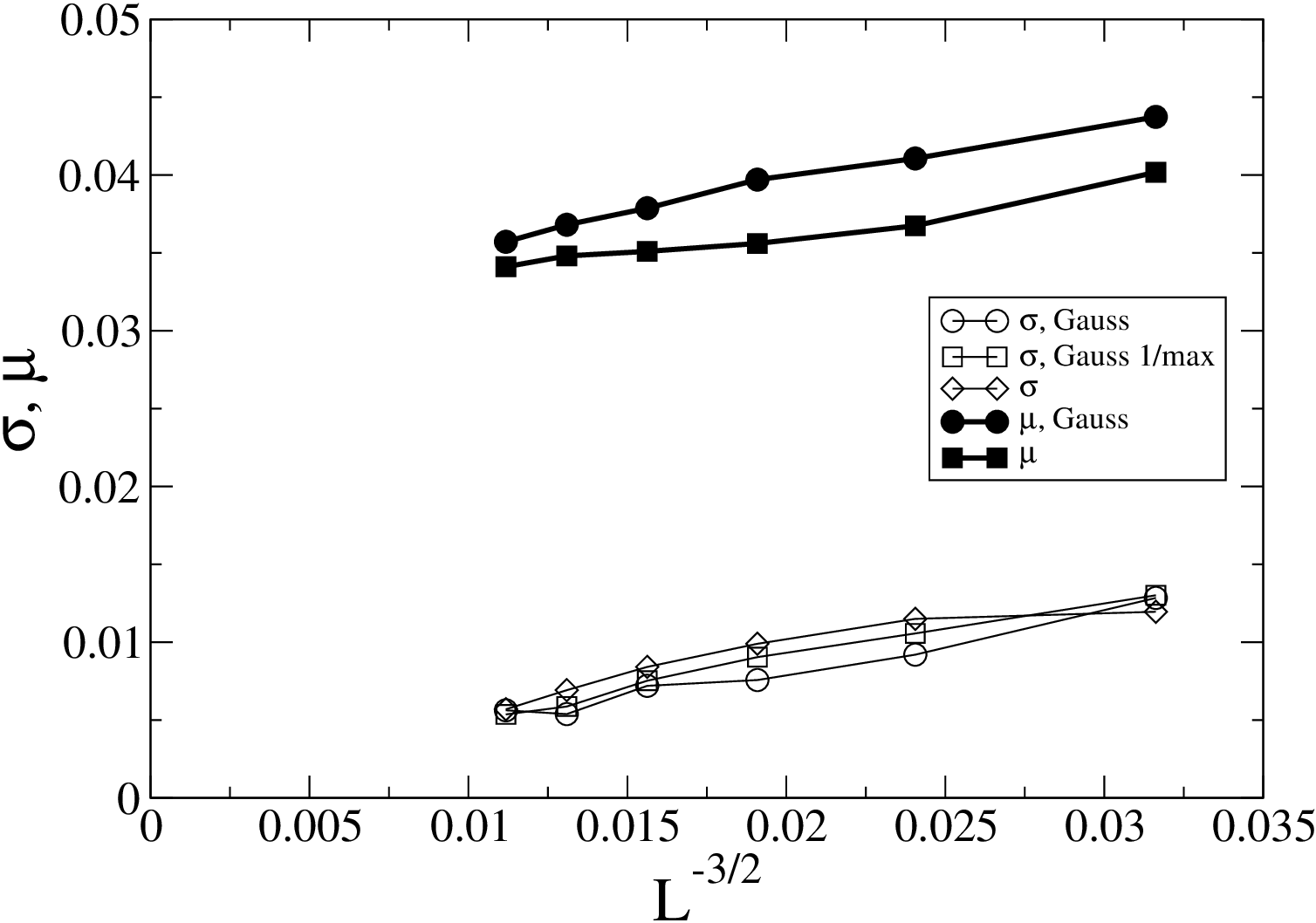}
\caption{The average mutual giant component, $\mu$, and its standard deviation $\sigma$, computed for $D=6$, $r=1$ as functions of the lattice size $L$ plotted against $N^{-1/4}=L^{-3/2}$. One can see approximately linear behavior for both quantities. Different curves correspond to different methods of estimating $\sigma$ and $\mu$. The first method is the direct computation of average $\mu$ and variance from realizations of $\mu>\mu_{min}$, where $\mu_{min}$ is the value of the minimum of the PDF. Another method is from the Gaussian fit near the maximum of the right peak of the PDF. In this case $\sigma$ can be computed from the maximum of the PDF and from the coefficient of the second power of the quadratic polynomial
fitting the logarithm of the PDF.}
\label{f:fit}
\end{figure}

Another example is the case of 6D lattice with $r=1$. For $L=10$, the transition looks second order (Fig. \ref{f:Lpc}). As we increase $L$, the transition begins to slowly shift from second order to first order. When $L=20$ the transition becomes distinctly first order. This demonstrates that the true type of transition for a 6D lattice at $r=1$ is first order. The explanation of this fact is based on the statistical errors in finite systems. If we remove exactly $N(1-p)$ random sites from the system in an initial attack, it does not mean that the size of the giant component in the lattice after the first stage of the cascade will be exactly $Ng(p)$, where $g(p)$ is the expected value of the giant component in a percolation problem. According to the law of large numbers the size of the giant component will be distributed around $g(p)$ with a standard deviation $\sigma_g\sim 1/\sqrt{N}$. Moreover, the long cascade of failures at $p=p_c$ can be viewed as sequence or iterations approaching the tangential point between the curve $y=pg(x)$ and $y=x$ \cite{Buldyrev2010}. If $g(x)$ is changed by an error $\sigma_g$, the root of the equation $x=pg(x)$ will change as $\sqrt{\sigma_g}$, because at the tangential point this equation becomes a quadratic equation with zero discriminant
and hence changes in discriminant of the order of $\sigma_g$ will result in the change of the root of the order of $\sqrt{\sigma_g}\sim N^{-1/4}$.
Thus, we can expect that the statistical error of the mutual giant component
as well as its mean value near the first order transition will decrease with the system size as $N^{-1/4}$. We observe this behavior for all $r$
and $D$. The PDF of the mutual giant component near the first order phase
transition is the derivative of the cumulative distribution and hence the inflection point of the plateau of the cumulative distribution corresponds to the minimum of the PDF. Thus PDF of $\mu$  near the first order phase transition is a bimodal distribution with a left peak corresponding to the collapsed states of the system, and a right peak corresponding to the survived states of the system. Figure \ref{f:bimodal} shows PDF of $\mu$ for $D=6, r=1$ for various values of $L$. Indeed one can see that the right peak becomes sharper as $L$ increases. Figure~\ref{f:fit} shows the standard deviation and mean of the right peak as function of $1/N^{-1/4}=1/L^{-3/2}$. Indeed, one can see an approximately linear behavior confirming our theory. As the system size increases, the right peak becomes narrower and for $L=20$ practically stops overlapping with the left peak, making the distribution first order-like.  


The upper critical dimension of the classical percolation might play an important role in the mutual
percolation problem with distance restriction as well. This means that, qualitatively, the behavior of our model for 
$D\geq 6$ should coincide with the behavior of a RR network with $k=2D=12$. In this RR network, the first value of $r$ in which there is a first order transition, is $r_I=1$ \cite{korn}. As shown above, when analyzing very large 6D lattice networks for 
which the finite-size effects become negligible (for $L\geq20$), the transition at $r=1$ is first order as well. For $D=7$ and $L=10$, the transition for $r=1$ is a clear first order transition. This supports our hypothesis that the upper critical dimension for percolation plays a role in the problem of mutual percolation with restricted interdependency distance. However, the quantitative difference of the behavior of $p_c$ for lattices and RR graphs gradually decreases with $D$.

\section{Conclusion}
In our study we confirm that the behavior of the interdependent $D$-dimensional lattices with distance limitation between the interdependent nodes, $r$, approaches the behavior of the interdependent RR graphs as $D$
increases. We find that for $D<6$ there is a value of $r=r_I>1$ such that for $r\geq r_I$ the cascading failures happen as a discontinuous first order transition, while for $r<r_I$ the transition is a continuous second order transition, as in the classical percolation theory. 

We also find
that in all dimensions, the interdependent lattices reach maximal vulnerability (largest $p_c$) at a distance $r=r_{max}>r_I$, such that for $r>r_{max}$ the vulnerability starts to decrease as $r\to\infty$. These findings are in qualitative agreement with Li {\it et al.}\cite{Li} who have found that for $D=2$, $r_I=r_{max}=8$. For $D=2$ we find $r_I=11$, $r_{max}=12$. The quantitative difference between our results can be explained by the fact that we use a shortest path as a metric while Li {\it et al.} use a maximal coordinate difference as a metric. The number of proximal nodes in Li {\it et al.} for $r=8$ is hence $(2r+1)^2=289$ while in our model the number of proximal nodes for $r=11$ is $1+2r(r+1)=265$ and for $r=12$ it is 313. Thus in terms of number of proximal nodes the value found Li {\it et al} for $r_I=r_{max}=8$ falls exactly in between our values $r_I=11$ and $r_{max}=12$. 

Note that as $D$ increases, both $r_I$ and $r_{max}$ decrease, but their difference increases. Moreover, the difference between $p_c[D,r_{max}]$ and $p_c[D,\infty]$ decreases exponentially with $D$.

More significantly we find that for $D=6$ and $r=1$, the transition is first order. This coincides with RR graphs with $r=1$ and large $k>8$.
This finding suggests that the upper critical dimension of the classical percolation, 
$D=6$, plays an important role in the problem of mutual percolation with distance restrictions.

We also investigate how the nature of the transition change as number of lattice sites $N\to\infty$. We find that when $N$ increases, the value of $p_c$ near the maximum increases logarithmically with $N$, approaching the value of $p_c^f$, the depinning transition of the propagation of the hole perimeter.The problem of the upper critical dimension for this depinning transition and its universality class is an interesting problem, which requires further investigation. $r_I$ and $r_{max}$ have a tendency to increase with $N$. However, this dependence is small and can be observed only for $D=2$.  

We also discover that close to $r_I$, the true order of the transition in the thermodynamic limit can be identified only for very large $N$. The bimodality of the distribution of the giant component indicated by the inflection point in the cumulative distribution, may either disappear, suggesting that the true nature of the transition for $N\to \infty$ is second order, or can become stronger, indicating that the transition is first order in the thermodynamic limit.  

\section{ACKNOWLEDGMENTS}
We wish to thank DTRA \#HDTRA1-14-1-0017 for financial support and Shlomo
Havlin for stimulating discussions. We acknowledge the partial
support of this research through the Dr. Bernard W. Gamson
Computational Science Center at Yeshiva College.


\begin{thebibliography}{99}

\bibitem{Albert2002}
R. Albert, R. and A. L. Barab\'{a}si, Rev. Mod. Phys. {\bf 74}, 47-97 (2002).

\bibitem{Dorogovtsev2003}
S. N. Dorogovtsev and J. F. F. Mendes, {\it Evolution of Networks: From Biological Nets to the Internet and WWW (Physics)} (Oxford Univ. Press, 2003).

\bibitem{Satorras2004}
R. P. Satorras and A. Vespignani, {\it Evolution and Structure of the Internet: A Statistical Physics Approach} (Cambridge Univ. Press, 2004).

\bibitem{Newman2006}
M. E. J. Newman, A.-L. Barab\'{a}si and D. J. Watts, {\it The Structure and Dynamics of Networks} (Princeton Univ. Press, 2006).

\bibitem{Caldarelli2007}
G. Caldarelli, and A. Vespignani, {\it Large Scale Structure and Dynamics of Complex Webs} (World Scientific, 2007).

\bibitem{Newman2010}M. E. J. Newman, {\it Networks: An Introduction} (Oxford Univ. Press, 2010).

\bibitem{Cohen2010}
R. Cohen and S. Havlin, {\it Complex Networks: Structure, Robustness, and Function} (Cambridge Univ. Press, 2010).

\bibitem{Buldyrev2010}
S. V. Buldyrev, R. Parshani, G. Paul, H. E. Stanley, and S. Havlin, Nature {\bf 464}, 1025-1028 (2010)

\bibitem{Parshani2010}
R. Parshani, S. V. Buldyrev, and S. Havlin, Phys. Rev. Lett. {\bf 105}, 048701 (2010)


\bibitem{Buldyrev2011}
S. V. Buldyrev, N. W. Shere, and G. A. Cwilich, Phys. Rev. E {\bf 83}, 016112 (2011).
\bibitem{Gao2012}
J. Gao, S. V. Buldyrev, H. E. Stanley, and S. Havlin, Nature Physics {\bf 8}, 40-48 (2012).

\bibitem{Son2011}
S. W. Son, P. Grassberger, and M. Paczuski, Phys. Rev. Lett.
{\bf 107}, 195702 (2011).

\bibitem{Son2012}
S. W. Son, G. Bizhani, C. Christensen, P. Grassberger, and
M. Paczuski, Europhys. Lett. {\bf 97}, 16006 (2012)

\bibitem{Baxter2012}
G. J. Baxter, S. N. Dorogovtsev, A. V. Goltsev, and J. F. F.
Mendes, Phys. Rev. Lett. {\bf 109}, 248701 (2012).

\bibitem{Peixoto2012}
T. P. Peixoto and S. Bornholdt, Phys. Rev. Lett. {\bf 109}, 118703
(2012).

\bibitem{Li} W. Li, A. Bashan, S. V. Buldyrev, H. E. Stanley and S. Havlin, Phys. Rev. Lett. \textbf{108}, 228702 (2012).

\bibitem{Bashan}
A. Bashan, Y. Berezin, S. V. Buldyrev, and S. Havlin, Nature Physics 9, 667-672 (2013).

\bibitem{korn} Y. Kornbluth, S. Lowinger, G. Cwilich and S.V. Buldyrev, Phys. Rev. E \textbf{89}, 032808 (2014).

\bibitem{Stauffer1994} D. Stauffer and A. Aharony, \textit{Introduction to Percolation Theory}, 2nd ed (Taylor and Francis, London, 1994).

\bibitem{Bunde1996}
\textit{Fractals and Disordered Systems}, 2nd ed., edited by A. Bunde
and S. Havlin (Springer, Berlin, 1996).

\bibitem{Barabasi1995} A.-L. Barabasi and H. E. Stanley, 
''Fractal concepts in surface growth'', (Cambridge University Press, New York, 1995).

\bibitem{Coniglio1984}
A. Coniglio, in {\it Finely Divided Matter}, Proceedings of the Les Houches Winter Conference, edited by N. Boccara and M. Daud (Springer Verlag, New York, 19	85); H.J. Hermann et al. J Phys A {\bf 17} L261 (1984).

\end{thebibliography}
\end{document}